\documentclass[11pt]{article}
\pdfoutput=1

\usepackage{amsmath,amsthm}
\usepackage{preprint}

\usepackage[utf8]{inputenc}
\usepackage[T1]{fontenc}
\usepackage{graphicx}
\usepackage{float}
\usepackage{booktabs}
\usepackage{array}
\newcolumntype{P}[1]{>{\raggedright\arraybackslash}p{#1}}
\usepackage{hyperref}
\usepackage{cleveref}
\usepackage{xcolor}
\usepackage{tikz}
\usetikzlibrary{arrows.meta, positioning, shapes.geometric, fit, calc}
\usepackage{listings}
\usepackage{enumitem}
\usepackage{caption}
\usepackage{subcaption}
\usepackage{needspace}

\lstdefinelanguage{Isabelle}{
  morekeywords={locale, fixes, assumes, shows, lemma, theorem, definition,
    interpretation, datatype, fun, where, record, type_synonym,
    by, auto, cases, metis, unfold_locales, rule, proof, qed,
    Some, None, let, in, case, of, begin, end, section, text,
    imports, theory, corollary, using, from, obtain, then, have,
    show, with, next, also, finally, hence, thus, moreover,
    assumes, and, not, if, else, True, False},
  sensitive=true,
  morecomment=[s]{(*}{*)},
  morestring=[b]",
  literate=
    {⇒}{{$\Rightarrow$}}1
    {⟹}{{$\Longrightarrow$}}1
    {⟷}{{$\longleftrightarrow$}}1
    {⟦}{{$[\![$}}1
    {⟧}{{$]\!]$}}1
    {∈}{{$\in$}}1
    {∉}{{$\notin$}}1
    {⊆}{{$\subseteq$}}1
    {∀}{{$\forall$}}1
    {∃}{{$\exists$}}1
    {∧}{{$\wedge$}}1
    {¬}{{$\neg$}}1
    {≡}{{$\equiv$}}1
    {≤}{{$\leq$}}1
    {≥}{{$\geq$}}1
    {≠}{{$\neq$}}1
    {λ}{{$\lambda$}}1
    {⊥}{{$\bot$}}1
    {×}{{$\times$}}1
    {∘}{{$\circ$}}1
    {⟶}{{$\longrightarrow$}}1
}

\newtheorem{theorem}{Theorem}[section]

\newtheorem{property}[theorem]{Property}

\hypersetup{
  colorlinks=true,
  linkcolor=blue!70!black,
  citecolor=blue!70!black,
  urlcolor=blue!70!black,
}

\newcommand{\isa}[1]{\lstinline[language=Isabelle]{#1}}
\newcommand{\isaf}[1]{\texttt{#1}}

\title{The Cross-Domain State Preservation Functor: A Mechanized Theory of Regulatory State Synchronization in Isabelle/HOL}

\author{
  Jinwook Kim\textsuperscript{1,2}~\href{https://orcid.org/0009-0004-4993-8005}{\mbox{\scalerel*{\begin{tikzpicture}[yscale=-1,transform shape]\pic{orcidlogo};\end{tikzpicture}}{|}}} \\
  (for the Oraclizer Core Team) \\
  \textsuperscript{1}Oraclizer Labs, Delaware, USA \\
  \textsuperscript{2}Oraclizer Labs Korea, Seoul, Korea \\
  \texttt{jay@oraclizer.io}
}

\date{}

\begin{document}

\maketitle

\begin{abstract}
A regulatory action on a tokenized asset---a freeze, a seizure, a confiscation---must take effect on
every domain holding it, or on none. We mechanize cross-domain state preservation in
Isabelle/HOL as a \emph{functor}: state machines are objects, structure-preserving synchronization
maps are morphisms, and identity, composition and associativity hold as theorems. Four results
follow. \emph{Safety}: a regulatory transition is reflected across all connected domains, with
roundtrip preservation, $N$-domain consistency, per-asset isolation and preserved terminal states,
so regulatory finality survives. \emph{Liveness}: deterministic conflict resolution
and starvation freedom under $f < n/3$ Byzantine nodes---a cardinality-bounded dishonest tag,
not a message-level adversary---and a fair-leader assumption, with $n \geq 3f+1$ shown to make
that fairness assumption inhabitable. \emph{Convergence}: from an arbitrary
unlocked configuration, with no initial consistency assumed, synchronization reaches a valid state
in boundedly many steps, along a recovery path that neither manufactures nor erases confiscations.
\emph{Hierarchy}: a tower of synchronization-degree functors linked by natural transformations
closed under composition---a layer that, to our knowledge, Lochbihler and Mari\'{c}'s
\isaf{ADS\_Functor} does not develop---with a one-directional degree monotonicity. We couple the
functor to that authenticated data structure, instantiated on a recursive Canton
transaction-tree model with a declared consensus-scope limit. Eighteen interpretations, an
external-domain instance and deletion-sensitive witnesses keep the theorems off the empty class. The
model is atomic and its refinement to code is unproven. Ten theory files build without \isa{sorry}
or \isa{oops}, released as a versioned Isabelle session.
\end{abstract}

\section{Introduction}
\label{sec:introduction}

Guaranteeing the consistency of regulatory compliance states is a fundamental challenge when tokenized assets operate across multiple blockchain networks and off-chain ledgers. When an asset freeze order is issued in one jurisdiction, it must be reflected immediately, deterministically, and with structural preservation across every domain where that asset exists. Without this guarantee, regulatory arbitrage becomes structurally possible.

This problem involves two independent dimensions. The first is \emph{safety}: if synchronization is performed correctly, is the resulting state consistent across all domains? The second is \emph{liveness}: in a decentralized environment with Byzantine nodes, does synchronization actually proceed---that is, does the system avoid halting, and is no pending regulatory request delayed indefinitely?

To the best of our knowledge, no prior work has formally verified both properties simultaneously for cross-domain regulatory state synchronization. Lochbihler and Mari\'{c}'s Merkle Functor pattern~\cite{lochbihler2020} formalized authenticated data structures within the Canton protocol in Isabelle/HOL, but its scope is limited to unidirectional data integrity within a single domain; it addresses neither cross-domain state preservation nor Byzantine environments. Velisarios~\cite{rahli2018} verified PBFT's safety in Coq, but liveness was out of scope. Byzantine consensus liveness has been verified across a range of tools: threshold-automata model checking~\cite{konnov2018} has been applied holistically to the safety and liveness of the Red Belly Blockchain consensus for any $f < n/3$~\cite{bertrand2022}, Wanner et al.\ verified a Byzantine log-replication protocol in Isabelle/HOL using the Heard-Of model~\cite{wanner2020}, and Losa and Dodds verified safety and liveness of the Stellar Consensus Protocol in a combination of Ivy and Isabelle/HOL~\cite{losa2020}. To the best of our knowledge, however, no prior work has used Isabelle/HOL alone to verify BFT consensus liveness in a blockchain context, and none in any tool has formally verified regulatory-specific consensus properties such as priority determinism and regulatory request starvation freedom. Nor, to our knowledge, has cross-domain state preservation been organized as a functor, equipped with an explicit synchronization-degree hierarchy, and coupled to an authenticated data structure across domains; that structural development is the core of this paper, with safety and liveness as the substrate it organizes.

Our contributions are as follows:

\begin{enumerate}[leftmargin=*]
\item \textbf{The cross-domain state-preservation functor.} We prove the category laws (identity, composition, associativity) as theorems for structure-preserving synchronization maps between heterogeneous domains, and read each transition system as a functor on the free monoid of action words. To the best of our knowledge, this functorial structure has not previously been mechanized across domains.

\item \textbf{Bidirectional, multi-domain state preservation (safety).} We extend the single-domain Merkle Functor pattern to bidirectional roundtrip guarantees, $N$-domain consistency, and per-asset isolation, unified in a single locale hierarchy; terminal states are preserved along every preservation morphism, so regulatory finality survives synchronization.

\item \textbf{Regulatory consensus liveness under Byzantine faults.} We verify deterministic resolution of conflicting regulatory actions and starvation freedom under $f < n/3$ Byzantine faults, with the threshold $n \geq 3f+1$ shown to be load-bearing for an inhabitable liveness locale. To the best of our knowledge, this is the first use of Isabelle/HOL alone to verify regulatory-specific consensus liveness.

\item \textbf{Guarded bounded convergence.} We prove that from an arbitrary unlocked configuration, with no assumption that cross-chain consistency holds initially, synchronization converges to a valid state within a bounded number of steps under a fair-leader assumption, driven by a well-founded measure on cross-chain inconsistency; the recovery path is terminal-faithful, neither manufacturing a confiscation that no chain held nor erasing one that a chain did hold.

\item \textbf{The synchronization-degree functor tower.} We build a tower of degree-indexed functors connected by natural transformations closed under composition---a layer that, to our knowledge, \isaf{ADS\_Functor} does not develop---together with a genuinely one-directional degree monotonicity.

\item \textbf{Coupling to an authenticated data structure.} We lift the merge and blinding operations of Lochbihler and Mari\'{c}'s \isaf{ADS\_Functor} to the global-state level and instantiate the coupling on the blindable-position functor and on a recursive model of the Canton transaction tree, with an explicitly declared consensus-scope limit.

\item \textbf{A reusable proof architecture.} The development is organized as four stacked layers---a generic locale layer, an instance layer, a compositional layer, and an authenticated layer---with ten domain-independent generic locales, Eisbach methods that discharge whole instances of the two entry-point locales by search, and an instance outside the regulatory domain that imports none of the regulatory theories. \Cref{sec:architecture} states what each layer fixes, what it costs an importer, and what the automation does not reach; parameter constraints derived from the proofs ($n \geq 3f + 1$, a fairness bound) translate into implementation design rationale.

\item \textbf{An application-domain contribution, stated as such.} The object of the formalization is the atomicity of \emph{regulatory effect} across heterogeneous domains: that a freeze, a seizure or a confiscation lands everywhere or nowhere, and that a confiscation, once recorded, is neither invented nor erased by the machinery that restores consistency. To the best of our knowledge, that object has not previously been mechanized; \Cref{sec:positioning} separates this application claim from the methodological ones so that each can be judged on its own evidence.
\end{enumerate}

The application context is a regulatory state transition model based on the RCP (Regulatory Compliance Protocol) framework (arXiv:2603.29278)~\cite{kim2026rcp}, which systematizes 31 requirements from 15 global financial regulatory authorities into five principles; the five regulatory states and seven regulatory actions formalized here are distilled in this work from the framework's enforcement requirements. This paper introduces the regulatory state transition model self-contained; detailed regulatory justifications are deferred to~\cite{kim2026rcp}.

The development comprises ten Isabelle/HOL theory files, building without \isa{sorry} or \isa{oops}: four base theories for the safety and liveness substrate, and six further theories developing the functor and its category laws, guarded bounded convergence, the synchronization-degree tower, the authenticated-data-structure coupling, an external instance, and a reusable proof-automation layer. The development is released publicly as a versioned Isabelle session with a build manifest recording the tool version, the source tree hash, and the build contract~\cite{kim2026artifact}; all artifacts are available on GitHub~\cite{oraclizer2026github}.

The remainder of this paper is structured as follows. \Cref{sec:model} defines the system and threat models and \Cref{sec:regulatory} the regulatory state transition model. \Cref{sec:safety} presents cross-domain state preservation (safety) and \Cref{sec:functor} promotes it to a functor. \Cref{sec:liveness} presents liveness under Byzantine faults and \Cref{sec:combined} the composition and bounded convergence of the two properties. \Cref{sec:tower} develops the synchronization-degree functor tower and \Cref{sec:ads} the coupling to Canton's authenticated data structure. \Cref{sec:genericity} sets out the layered proof architecture, the automation and the non-vacuity evidence, and closes with the lessons that generalize past this subject matter. \Cref{sec:related} surveys related work and positions the application and methodological claims separately, \Cref{sec:discussion} states the scope boundary, and \Cref{sec:conclusion} concludes.

\section{System Model and Threat Model}
\label{sec:model}

\subsection{System Model}
\label{sec:system-model}

The formal model comprises the following elements.

\paragraph{Domains.} A finite set of domains (chains) $\mathcal{D}$. Each domain shares the same state machine structure. A domain abstracts a blockchain network, an off-chain ledger, or any system supporting state transitions.

\paragraph{State transitions.} A deterministic partial function:
\begin{equation}
\label{eq:transition}
\delta : S \times A \rightarrow S_\bot
\end{equation}
where $S$ is a finite state set, $A$ is a finite action set, and $S_\bot = S \cup \{\bot\}$. The same (state, action) pair always yields the same result. Terminal states $T \subseteq S$ return $\bot$ for all transitions.

\paragraph{Per-asset locking.} Preemptive locking at the \isaf{asset\_id} granularity. Lock acquisition and release are atomic. Each synchronization operation holds at most one lock on one asset (single-resource locking).

\paragraph{Synchronization.} The \isaf{sync} function executes atomically: (1)~verify the asset exists on the source chain, (2)~validate the transition, (3)~acquire a lock, (4)~update all connected chains, (5)~release the lock. If any step fails, the entire operation returns \isaf{None}.

\paragraph{Communication.} The safety proof (Property~1) assumes reliable channels. Message loss, duplication, and reordering are outside the model. Extending to a partially synchronous network model remains an open problem.

\paragraph{Global state.} Formalized in Isabelle/HOL as:

\begin{lstlisting}
record global_state =
  gs_chains :: "chain_id ⇒ chain_state"
  gs_locks  :: "asset_id ⇒ bool"
\end{lstlisting}

\noindent where \isaf{chain\_state = asset\_id $\Rightarrow$ asset\_state option} and \isaf{asset\_state} is a record containing the asset identifier, regulatory state, owner, and lock status.

\paragraph{Global validity.} Defined as the conjunction of two conditions:

\begin{lstlisting}
definition valid_state :: "global_state ⇒ bool" where
  "valid_state gs ≡ consistent_state gs ∧ no_locked_without_reason gs"
\end{lstlisting}

\noindent \isaf{consistent\_state} requires that all chains holding the same asset agree on its regulatory state. \isaf{no\_locked\_without\_reason} requires that no asset is locked in quiescent states.

\subsection{Threat Model}
\label{sec:threat-model}

\paragraph{Property~1 (Safety).} All nodes follow the protocol honestly. Concurrent regulatory actions are serialized via preemptive locking. Network delays and Byzantine behavior are out of scope.

\paragraph{Property~2 (Liveness).} The honest-node assumption of Property~1 is relaxed. Up to $f$ nodes may exhibit arbitrary malicious behavior (Byzantine faults), where $n \geq 3f + 1$. Byzantine leaders may be elected consecutively; a fairness assumption bounds the duration. Byzantine behaviour itself is abstracted: the model represents Byzantine nodes as a cardinality-bounded dishonest tag whose epochs carry no progress guarantee; adversarial message-level actions (equivocation, forgery) and the underlying network messaging are not formalized (\Cref{sec:assume-guarantee}). Under the atomic synchronization model there is no concurrent lock contention, so forced lock release under contention is out of scope.

\paragraph{Combined.} Property~2 establishes liveness (deterministic resolution, starvation freedom) in the Byzantine environment. It does not by itself discharge Property~1's honest-execution premise. The genuine safety--liveness fusion is a separate bounded-convergence result of the compositional layer (\Cref{sec:combined}): from an arbitrary unlocked initial state, with no assumption that cross-chain consistency holds initially, synchronization converges to a valid state within a bounded number of steps under the fair-leader assumption.

\subsection{Non-Goals}
\label{sec:non-goals}

We explicitly state what this model does \emph{not} guarantee:

\begin{itemize}[leftmargin=*]
\item \textbf{Chain-specific characteristics}: gas costs, block times, etc.\ are fully abstracted.
\item \textbf{Regulatory authority verification}: whether a specific authority has permission to execute a given action is not verified.
\item \textbf{Financial operations}: balances, token transfers, etc.\ are outside the model scope.
\item \textbf{Double-spend prevention}: this model addresses state consistency for concurrent regulatory actions only.
\item \textbf{Refinement}: formal correspondence between the model and Rust/Solidity code is unproven.
\item \textbf{Open systems}: dynamic arrival of new requests is not modeled (\isaf{non\_honest\_bounded} assumption).
\end{itemize}

\section{Regulatory State Transition Model}
\label{sec:regulatory}

This section introduces the regulatory state transition model that is the subject of formal verification. The model comprises five regulatory states and seven regulatory actions, distilled in this work from the enforcement requirements of the RCP framework~\cite{kim2026rcp}, which systematizes 31 requirements from 15 global financial regulatory authorities into five principles. Detailed regulatory justifications are available in~\cite{kim2026rcp}.

\subsection{States and Actions}
\label{sec:states-actions}

Five regulatory states represent the regulatory positions a tokenized asset can occupy:

\begin{lstlisting}
datatype reg_state = ACTIVE | FROZEN | SEIZED | CONFISCATED | RESTRICTED
\end{lstlisting}

\begin{itemize}[leftmargin=*]
\item \textbf{ACTIVE}: normal trading permitted.
\item \textbf{FROZEN}: temporary trading suspension, reversible.
\item \textbf{SEIZED}: court-ordered custody; ownership retained but control lost.
\item \textbf{CONFISCATED}: permanent ownership revocation (terminal state).
\item \textbf{RESTRICTED}: trading permitted under specific conditions only.
\end{itemize}

Seven regulatory actions trigger transitions between states:

\begin{lstlisting}
datatype reg_action = FREEZE | SEIZE | CONFISCATE | RESTRICT
                    | UNFREEZE | UNRESTRICT | RELEASE
\end{lstlisting}

\subsection{Transition Function}
\label{sec:transition-function}

The transition function \isaf{reg\_transition} defines 35 (state, action) combinations, of which 12 are valid transitions and 23 return \isaf{None}. Since it is defined using Isabelle/HOL's \isa{fun} keyword, determinism is trivially guaranteed.

\begin{figure}[t]
\centering
\begin{tikzpicture}[
  state/.style={rectangle, rounded corners=3pt, draw=black, fill=gray!5, minimum width=2.2cm, minimum height=0.8cm, font=\small\sffamily},
  terminal/.style={state, fill=black!85, text=white, draw=black},
  trans/.style={-{Stealth[length=5pt]}, thick, color=blue!70!black},
  revtrans/.style={-{Stealth[length=5pt]}, thick, color=green!50!black, dashed},
  escalate/.style={-{Stealth[length=5pt]}, thick, color=red!70!black},
  lbl/.style={font=\footnotesize, inner sep=1.6pt, fill=white, fill opacity=0.9,
              text opacity=1, rounded corners=1pt}
]
  \node[state] (active) {ACTIVE};
  \node[state, above right=2.0cm and 3.1cm of active] (frozen) {FROZEN};
  \node[state, below right=2.0cm and 3.1cm of active] (restricted) {RESTRICTED};
  \node[state, right=6.2cm of active] (seized) {SEIZED};
  \node[terminal, right=3.1cm of seized] (confiscated) {CONFISCATED};

  \draw[trans] (active) to[bend left=24] node[lbl, sloped, above] {FREEZE} (frozen);
  \draw[trans] (active) to[bend right=24] node[lbl, sloped, below] {RESTRICT} (restricted);
  \draw[trans] (active) to[bend left=9] node[lbl, pos=0.42, above] {SEIZE} (seized);

  \draw[revtrans] (frozen) to[bend left=24] node[lbl, sloped, below] {UNFREEZE} (active);
  \draw[revtrans] (restricted) to[bend right=24] node[lbl, sloped, above] {UNRESTRICT} (active);
  \draw[revtrans] (seized) to[bend left=9] node[lbl, pos=0.42, below] {RELEASE} (active);

  \draw[escalate] (frozen) -- node[lbl, above] {SEIZE} (seized);
  \draw[escalate] (restricted) to[bend left=16] node[lbl, sloped, right, pos=0.5] {FREEZE} (frozen);

  \draw[escalate] (active) to[bend left=34] node[lbl, pos=0.58, above] {CONFISCATE} (confiscated);
  \draw[escalate] (frozen) to[bend left=20] node[lbl, pos=0.5, above] {CONFISCATE} (confiscated);
  \draw[escalate] (seized) -- node[lbl, pos=0.5, above] {CONFISCATE} (confiscated);
  \draw[escalate] (restricted) to[bend right=34] node[lbl, pos=0.58, below] {CONFISCATE} (confiscated);

  \matrix[draw, below=1.5cm of active, anchor=north, font=\footnotesize,
          column sep=0.6em, row sep=2pt, inner sep=6pt] {
    \draw[trans] (0,0) -- (0.7,0); & \node[anchor=west]{Forward action}; \\
    \draw[revtrans] (0,0) -- (0.7,0); & \node[anchor=west]{Reversal}; \\
    \draw[escalate] (0,0) -- (0.7,0); & \node[anchor=west]{Escalation}; \\
  };
\end{tikzpicture}
\caption{Regulatory state transition diagram. Five states with 12 valid transitions. CONFISCATED (solid dark) is the terminal state: all transitions from it return \isaf{None}. CONFISCATE is universally reachable from every non-terminal state.}
\label{fig:state-diagram}
\end{figure}

\Cref{fig:state-diagram} depicts the transition diagram. The full transition matrix is shown in \Cref{tab:transition-matrix}.

\begin{table}[t]
\centering
\caption{Complete transition matrix. Target state is shown for valid transitions; \textnormal{--} denotes \isaf{None}.}
\label{tab:transition-matrix}
\footnotesize
\resizebox{\textwidth}{!}{%
\begin{tabular}{@{}l ccccccc@{}}
\toprule
& \textbf{Freeze} & \textbf{Seize} & \textbf{Confiscate} & \textbf{Restrict} & \textbf{Unfreeze} & \textbf{Unrestrict} & \textbf{Release} \\
\midrule
\textbf{Active}      & FROZEN & SEIZED & CONFISCATED & RESTRICTED & -- & -- & -- \\
\textbf{Frozen}      & -- & SEIZED & CONFISCATED & -- & ACTIVE & -- & -- \\
\textbf{Seized}      & -- & -- & CONFISCATED & -- & -- & -- & ACTIVE \\
\textbf{Confiscated} & -- & -- & -- & -- & -- & -- & -- \\
\textbf{Restricted}  & FROZEN & -- & CONFISCATED & -- & -- & ACTIVE & -- \\
\bottomrule
\end{tabular}%
}
\end{table}

\subsection{Design Decisions}
\label{sec:design-decisions}

Three design decisions are grounded in legal precedent and model simplification.

\paragraph{RECOVER/LIQUIDATE exclusion.} RECOVER (stolen asset return) and LIQUIDATE (forced liquidation), two of the product's six enforcement actions, are force-transfer or external-DEX operations rather than regulatory state transitions. This model formalizes the seven regulatory state-transition actions only.

\paragraph{SEIZED $\to$ FROZEN exclusion.} SEIZED imposes a strictly stronger legal constraint than FROZEN. ``Weakening'' a court-ordered seizure to a mere freeze is legally nonsensical. The path is: RELEASE $\to$ ACTIVE $\to$ FREEZE.

\paragraph{FROZEN $\to$ RESTRICTED exclusion.} These states arise from different regulatory contexts; transition requires returning to ACTIVE first: UNFREEZE $\to$ ACTIVE $\to$ RESTRICT.

\subsection{Fundamental Properties}
\label{sec:fundamental-properties}

Three fundamental properties are proven in Isabelle/HOL.

\begin{property}[Terminal Absorptivity (I1)]
CONFISCATED is terminal: all actions return \isaf{None}.
\end{property}

\begin{lstlisting}
lemma confiscated_terminal:
  "reg_transition CONFISCATED a = None"
  by (cases a) auto
\end{lstlisting}

\begin{property}[Universal Confiscation (I2)]
CONFISCATE is reachable from every non-terminal state.
\end{property}

\begin{lstlisting}
lemma confiscate_universal:
  "s ≠ CONFISCATED ⟹ reg_transition s CONFISCATE = Some CONFISCATED"
  by (cases s) auto
\end{lstlisting}

\begin{property}[No Self-Loops]
No valid transition returns to the same state.
\end{property}

\begin{lstlisting}
lemma no_self_loops:
  "reg_transition s a = Some s ⟹ False"
  by (cases s; cases a; auto)
\end{lstlisting}

Additionally, every non-terminal state admits at least one valid action (\isaf{non\_ter\-mi\-nal\_has\_action}), ensuring progress is always possible.

\section{Cross-Domain State Preservation}
\label{sec:safety}

\subsection{Reusable Locale Hierarchy}
\label{sec:locale-hierarchy}

The core contribution of Property~1 is a hierarchical abstraction comprising four generic locales, defined in \isaf{State\_Preservation.thy} (455~lines). These locales are domain-independent.

\paragraph{Notation in the listings.} Every displayed \isa{theorem}, \isa{lemma} and \isa{locale} in this paper is the statement as it appears in the sources, with two presentational elisions: locale headers omit the \isa{for} block that repeats the inherited parameters with their type annotations, and proof scripts are omitted. Assumption labels, hypotheses and conclusions are reproduced verbatim.

\begin{figure}[!b]
\centering
\begin{tikzpicture}[
  box/.style={rectangle, rounded corners=2pt, draw=black, fill=white, minimum width=4.2cm, minimum height=1.4cm, font=\small\sffamily, align=center},
  arr/.style={-{Stealth[length=5pt]}, thick},
  note/.style={font=\footnotesize\itshape, text=black!80, align=left},
  node distance=1.8cm
]
  \node[box] (sm) {\textbf{state\_machine}\\{\footnotesize finite states, actions,}\\{\footnotesize deterministic transition, terminal}};
  \node[box, below=of sm] (sp) {\textbf{state\_preservation}\\{\footnotesize naturality condition:}\\{\footnotesize sync commutes with transition}};
  \node[box, below=of sp] (ssp) {\textbf{symmetric\_state\_preservation}\\{\footnotesize inverse maps,}\\{\footnotesize roundtrip guarantee}};
  \node[box, below=of ssp] (mdp) {\textbf{multi\_domain\_preservation}\\{\footnotesize $N$ domains,}\\{\footnotesize consistency + isolation}};

  \draw[arr] (sm) -- (sp);
  \draw[arr] (sp) -- (ssp);
  \draw[arr] (ssp) -- (mdp);

  \node[note, right=1cm of sm] {apply\_actions\\terminal\_absorbing\\{\color{black!60}inst: reg\_sm}};
  \node[note, right=1cm of sp] {sequential\_preservation\\sequential\_preservation\_none\\{\color{black!60}inst: escalation, layer fwd/bwd}};
  \node[note, right=1cm of ssp] {state\_map\_injective\\(no information loss)\\{\color{black!60}inst: onchain DAML bridge}};
  \node[note, right=1cm of mdp] {cross\_domain\_consistency\\sync\_isolation\\{\color{black!60}inst: parametric (finite doms)}};
\end{tikzpicture}
\caption{Locale hierarchy for cross-domain state preservation (Property~1). Each level adds assumptions and proves additional properties; theorems proven at each level are shown on the right, with the concrete regulatory instance witnessing each locale (greyed). Every generic locale is instantiated, so none of the proven theorems holds only vacuously.}
\label{fig:locale-hierarchy}
\end{figure}

\subsubsection{state\_machine.}
The most basic locale defines a state machine with a finite state set, finite action set, deterministic partial transition function, and terminal states.

\needspace{16\baselineskip}
\begin{lstlisting}
locale state_machine =
  fixes states :: "'s set"
    and actions :: "'a set"
    and transition :: "'s ⇒ 'a ⇒ 's option"
    and terminal :: "'s set"
  assumes finite_states: "finite states"
    and finite_actions: "finite actions"
    and terminal_subset: "terminal ⊆ states"
    and terminal_absorbing:
      "⟦ s ∈ terminal; a ∈ actions ⟧ ⟹ transition s a = None"
    and transition_closed:
      "⟦ s ∈ states; a ∈ actions;
         transition s a = Some s' ⟧ ⟹ s' ∈ states"
    and transition_domain:
      "⟦ s ∉ states ⟧ ⟹ transition s a = None"
\end{lstlisting}

Within this locale, sequential action application (\isaf{apply\_actions}) and terminal absorptivity are proven.

\subsubsection{state\_preservation.}
Defines a structure-preserving map between two state machines. The key property is the \emph{naturality condition}: synchronization commutes with transition. Writing $\sigma$ for the state map \isaf{state\_map}, $\alpha$ for the action map \isaf{action\_map}, and $\delta_s, \delta_t$ for the source and target transition functions, naturality requires, for all $s \in S_s$ and $a \in A_s$,
\begin{equation}
\label{eq:naturality}
\delta_s(s, a) = \mathit{Some}\;s'
\;\;\Longrightarrow\;\;
\delta_t\bigl(\sigma(s),\, \alpha(a)\bigr) = \mathit{Some}\;\sigma(s'),
\end{equation}
together with the dual condition that $\delta_s(s,a) = \bot$ implies $\delta_t(\sigma(s), \alpha(a)) = \bot$ (failure is preserved). \Cref{eq:naturality} is the commutative square of \Cref{fig:naturality}: the two paths from $s$ to the target state space agree.

\begin{figure}[H]
\centering
\begin{tikzpicture}[
  node distance=2.6cm and 4.2cm,
  vert/.style={font=\normalsize},
  arr/.style={-{Stealth[length=5pt]}, thick},
  lbl/.style={font=\footnotesize, align=center}
]
  \node[vert] (s)   {$s$};
  \node[vert] (s2)  [right=of s]  {$s'$};
  \node[vert] (ts)  [below=of s]  {$\sigma(s)$};
  \node[vert] (ts2) [below=of s2] {$\sigma(s')$};

  \draw[arr] (s)  -- node[lbl, above] {$a$\\{\itshape\color{black!70}source $\delta_s$}} (s2);
  \draw[arr] (ts) -- node[lbl, below] {$\alpha(a)$\\{\itshape\color{black!70}target $\delta_t$}} (ts2);
  \draw[arr] (s)  -- node[lbl, left]  {$\sigma$} (ts);
  \draw[arr] (s2) -- node[lbl, right] {$\sigma$} (ts2);
\end{tikzpicture}
\caption{Naturality square for \isaf{state\_preservation}. The state map $\sigma$ (\isaf{state\_map}) commutes with transition: if the source transition $\delta_s$ maps $(s, a)$ to $\mathit{Some}\;s'$, then the target transition $\delta_t$ maps $(\sigma(s), \alpha(a))$ to $\mathit{Some}\;\sigma(s')$. This is the homomorphism condition that \Cref{sec:functor} promotes to a functor by proving the category laws, in the sense of Lochbihler and Mari\'{c}'s \isaf{ADS\_Functor}.}
\label{fig:naturality}
\end{figure}

\begin{lstlisting}
locale state_preservation =
  source: state_machine + target: state_machine +
  fixes state_map :: "'s ⇒ 't"
    and action_map :: "'a ⇒ 'b"
  assumes state_map_well_defined:
    "s ∈ states_s ⟹ state_map s ∈ states_t"
    and action_map_well_defined:
    "a ∈ actions_s ⟹ action_map a ∈ actions_t"
    and terminal_preservation:
    "s ∈ terminal_s ⟹ state_map s ∈ terminal_t"
    and naturality:
    "⟦ s ∈ states_s; a ∈ actions_s;
       transition_s s a = Some s' ⟧
     ⟹ transition_t (state_map s) (action_map a)
        = Some (state_map s')"
    and naturality_none:
    "⟦ s ∈ states_s; a ∈ actions_s;
       transition_s s a = None ⟧
     ⟹ transition_t (state_map s) (action_map a)
        = None"
\end{lstlisting}

The fundamental theorem extends naturality from single transitions to action sequences:

\begin{theorem}[Sequential Preservation]
\label{thm:sequential}
If an action sequence is valid at the source, the mapped sequence is valid at the target with the mapped final state.
\end{theorem}

\begin{lstlisting}
theorem sequential_preservation:
  assumes "s ∈ states_s"
    and "∀a ∈ set as. a ∈ actions_s"
    and "source.apply_actions s as = Some s'"
  shows "target.apply_actions (state_map s)
           (map action_map as) = Some (state_map s')"
\end{lstlisting}

The proof proceeds by structural induction on the action list. The base case (empty list) is trivial. The inductive step combines naturality for the first action with the induction hypothesis for the remainder. Failure preservation (\isaf{sequential\_preservation\_none}) follows the same structure.

\paragraph{Finality preservation.} Beyond naturality, the morphism preserves finality. \isaf{terminal\_preservation} sends terminal states to terminal states, and terminal absorptivity at the target then disables every action on the image of a terminal source state:

\begin{lstlisting}
corollary terminal_image_absorbing:
  assumes "s ∈ terminal_s" and "a ∈ actions_s"
  shows "transition_t (state_map s) (action_map a) = None"
\end{lstlisting}

In the regulatory instance the only terminal state is CONFISCATED (\Cref{sec:regulatory}), so this is the formal statement that a confiscation cannot be undone by carrying the asset to a domain on which some transition might still be enabled: regulatory finality survives synchronization, and the cross-domain escape route that motivates the model in \Cref{sec:introduction} is closed at the level of the preservation morphism itself.

\subsubsection{symmetric\_state\_preservation.}
For bidirectional synchronization, both directions must preserve structure. This locale adds inverse maps with roundtrip guarantees:

\begin{lstlisting}
locale symmetric_state_preservation =
  forward: state_preservation + backward: state_preservation +
  assumes roundtrip_state_src:
    "s ∈ states_s ⟹ state_map_inv (state_map s) = s"
    and roundtrip_state_tgt:
    "t ∈ states_t ⟹ state_map (state_map_inv t) = t"
    and roundtrip_action_src:
    "a ∈ actions_s ⟹ action_map_inv (action_map a) = a"
    and roundtrip_action_tgt:
    "b ∈ actions_t ⟹ action_map (action_map_inv b) = b"
\end{lstlisting}

From the roundtrip guarantee, injectivity of \isaf{state\_map} is derived, ensuring no information loss during synchronization:

\begin{lstlisting}
lemma state_map_injective:
  "⟦ s1 ∈ states_s; s2 ∈ states_s;
     state_map s1 = state_map s2 ⟧ ⟹ s1 = s2"
  using roundtrip_state_src by metis
\end{lstlisting}

\subsubsection{multi\_domain\_preservation.}
Generalizes to $N$ domains sharing the same abstract state machine. After synchronization on one domain, all connected domains reach the same resulting state.

\begin{lstlisting}
locale multi_domain_preservation =
  fixes domains :: "'d set"
    and states :: "'s set"
    and actions :: "'a set"
    and transition :: "'s ⇒ 'a ⇒ 's option"
    and terminal :: "'s set"
    and domain_state :: "'d ⇒ 'id ⇒ 's option"
  assumes fin_domains: "finite domains"
    and sm: "state_machine states actions
               transition terminal"
    and consistent_init:
      "⟦ d1 ∈ domains; d2 ∈ domains;
         domain_state d1 aid = Some s1;
         domain_state d2 aid = Some s2 ⟧
       ⟹ s1 = s2"
\end{lstlisting}

Two key theorems are proven:

\begin{theorem}[Cross-Domain Consistency]
\label{thm:consistency}
After \isaf{sync\_all}, every connected domain agrees on the new state.
\end{theorem}

\begin{lstlisting}
theorem cross_domain_consistency:
  assumes "source ∈ domains"
    and "domain_state source aid = Some s"
    and "s ∈ states"
    and "action ∈ actions"
    and "transition s action = Some s'"
    and "sync_all source action aid
           domain_state = Some ds'"
    and "d ∈ connected_domains aid"
  shows "ds' d aid = Some s'"
\end{lstlisting}

\begin{theorem}[Sync Isolation]
\label{thm:isolation}
Synchronization on one asset does not affect other assets.
\end{theorem}

\begin{lstlisting}
theorem sync_isolation:
  assumes "sync_all source action aid
             domain_state = Some ds'"
    and "aid' ≠ aid"
  shows "ds' d aid' = domain_state d aid'"
\end{lstlisting}

\subsection{Regulatory Instance}
\label{sec:regulatory-instance}

\isaf{Regulatory\_Instance.thy} (1,598~lines) instantiates the generic locales with the regulatory domain.

\paragraph{State machine instantiation.} We prove that \isaf{reg\_state}, \isaf{reg\_action}, and \isaf{reg\_transition} satisfy all assumptions of the \isaf{state\_machine} locale:

\begin{lstlisting}
interpretation reg_sm: state_machine
  reg_states reg_actions reg_transition reg_terminal
\end{lstlisting}

\paragraph{Synchronization protocol.} The \isaf{sync} function is defined in a lock--validate--update--unlock pattern (\Cref{fig:sync-workflow}):

\begin{figure}[t]
\centering
\begin{tikzpicture}[
  syncstep/.style={rectangle, rounded corners=2pt, draw=black, fill=blue!5, minimum width=2.8cm, minimum height=0.7cm, font=\small\sffamily},
  fail/.style={rectangle, rounded corners=2pt, draw=red!50, fill=red!5, minimum width=1.4cm, minimum height=0.5cm, font=\footnotesize\sffamily},
  arr/.style={-{Stealth[length=5pt]}, thick},
  farr/.style={-{Stealth[length=4pt]}, dashed, color=red!60},
  node distance=0.8cm
]
  \node[syncstep] (s1) {1. Check state};
  \node[syncstep, below=of s1] (s2) {2. Validate transition};
  \node[syncstep, below=of s2] (s3) {3. Acquire lock};
  \node[syncstep, below=of s3] (s4) {4. Update all chains};
  \node[syncstep, below=of s4] (s5) {5. Release lock};

  \node[fail, right=1.5cm of s1] (f1) {None};
  \node[fail, right=1.5cm of s2] (f2) {None};
  \node[fail, right=1.5cm of s3] (f3) {None};

  \draw[arr] (s1) -- (s2);
  \draw[arr] (s2) -- (s3);
  \draw[arr] (s3) -- (s4);
  \draw[arr] (s4) -- (s5);

  \draw[farr] (s1) -- node[above, font=\footnotesize, color=red!70!black] {not found} (f1);
  \draw[farr] (s2) -- node[above, font=\footnotesize, color=red!70!black] {invalid} (f2);
  \draw[farr] (s3) -- node[above, font=\footnotesize, color=red!70!black] {locked} (f3);

  \node[below=0.3cm of s5, font=\small\sffamily] {Some gs'};
\end{tikzpicture}
\caption{Synchronization protocol workflow. Solid arrows: normal execution. Dashed red arrows: failure paths returning \isaf{None}. The lock is acquired \emph{before} updates and released \emph{after}, ensuring atomicity.}
\label{fig:sync-workflow}
\end{figure}

\needspace{18\baselineskip}
\begin{lstlisting}
definition sync ::
  "chain_id ⇒ reg_action ⇒ asset_id
   ⇒ global_state ⇒ global_state option"
where
  "sync source action aid gs =
    (case get_reg_state gs source aid of
       None ⇒ None
     | Some current_st ⇒
         (case reg_transition current_st action of
            None ⇒ None
          | Some new_st ⇒
              (case acquire_lock gs aid of
                 None ⇒ None
               | Some gs_locked ⇒
                   let targets = connected_chains gs aid
                   in let gs_updated =
                     update_all_chains gs_locked aid
                       new_st targets
                   in Some (release_lock gs_updated
                              aid))))"
\end{lstlisting}

\paragraph{Regulatory homomorphism.} If \isaf{sync} succeeds, every connected chain agrees on the new regulatory state. The statement carries no validity or existence precondition: success of \isaf{sync} is itself the hypothesis, and the agreement of the connected chains follows from the definition of the update alone.

\begin{theorem}[Regulatory Homomorphism]
\label{thm:homomorphism}
\end{theorem}

\begin{lstlisting}
theorem regulatory_homomorphism:
  assumes "get_reg_state gs source aid = Some s"
    and "reg_transition s action = Some s'"
    and "sync source action aid gs = Some gs'"
    and "c ∈ connected_chains gs aid"
  shows "get_reg_state gs' c aid = Some s'"
\end{lstlisting}

The proof decomposes \isaf{sync} into lock, \isaf{update\_all\_chains}, and \isaf{release\_lock}, then shows: (1)~\isaf{acquire\_lock} does not change chain state, (2)~the asset exists on chain~$c$, (3)~\isaf{update\_all\_chains} correctly updates chain~$c$, and (4)~\isaf{release\_lock} does not change chain state.

\paragraph{Valid state preservation.} The \isaf{sync} operation preserves the \isaf{valid\_state} invariant, closing the inductive invariant:

\begin{theorem}[Valid State Preservation]
\label{thm:valid-state}
\end{theorem}

\begin{lstlisting}
theorem valid_state_preservation:
  assumes "valid_state gs"
    and "get_reg_state gs source aid = Some s"
    and "reg_transition s action = Some s'"
    and "sync source action aid gs = Some gs'"
  shows "valid_state gs'"
\end{lstlisting}

The proof decomposes into two parts: (1)~\isaf{consistent\_state} preservation: \isaf{sync} updates all connected chains to the same new state and leaves other assets untouched; (2)~\isaf{no\_locked\_without\_reason} preservation: \isaf{sync} acquires and releases the lock within the same operation.

\paragraph{No partial enforcement.} Together these theorems close the intermediate states that motivate the model. If \isaf{sync} succeeds, \Cref{thm:homomorphism} puts every connected chain at the new regulatory state and \Cref{thm:valid-state} keeps the global invariant; if any step fails (asset not found, invalid transition, contended lock), \isaf{sync} returns \isaf{None} and, being a pure function of the global state, changes nothing. Within the atomic model, a configuration in which a freeze has taken effect on some but not all of an asset's connected chains is therefore unreachable by the synchronization protocol: enforcement is all-or-nothing at the granularity of the connected-chain set. Partial propagation inside a non-atomic network is exactly the scope limit of \Cref{sec:system-model}, revisited in \Cref{sec:future-work}.

\paragraph{Multi-domain instantiation.} For any finite domain set and valid global state, all theorems of the generic locale apply to the regulatory model:

\begin{lstlisting}
theorem reg_multi_domain_instantiation:
  assumes "finite (doms :: chain_id set)"
    and "valid_state gs"
  shows "multi_domain_preservation doms reg_states
           reg_actions reg_transition reg_terminal
           (reg_domain_state gs)"
\end{lstlisting}

\subsection{Heterogeneous-Action Instance: Escalation Preservation}
\label{sec:escalation-instance}

The \isaf{state\_machine} and \isaf{multi\_domain\_preservation} locales are instantiated above; the \isaf{state\_preservation} and \isaf{symmetric\_state\_preservation} locales of \Cref{sec:locale-hierarchy} must also be instantiated with concrete witnesses, lest their theorems hold only vacuously. A locale whose theorems are never instantiated is satisfied trivially, so each abstraction in the hierarchy is exercised against a concrete cross-domain mapping. This subsection and the next provide those witnesses.

The first models two chains that share the regulatory state space but differ in their on-chain action vocabularies. Some jurisdictions process de-escalation (UNFREEZE, UNRESTRICT, RELEASE) exclusively through off-chain judicial procedures rather than as on-chain actions; the corresponding chain therefore exposes only the four escalation actions on chain. We capture this asymmetry with a distinct target action type:

\begin{lstlisting}
datatype chain_b_action =
  B_FREEZE | B_SEIZE | B_CONFISCATE | B_RESTRICT

definition escalation_actions :: "reg_action set" where
  "escalation_actions = {FREEZE, SEIZE, CONFISCATE, RESTRICT}"

fun chain_b_transition ::
  "reg_state ⇒ chain_b_action ⇒ reg_state option" where
  "chain_b_transition s B_FREEZE     = reg_transition s FREEZE"
| "chain_b_transition s B_SEIZE      = reg_transition s SEIZE"
| "chain_b_transition s B_CONFISCATE = reg_transition s CONFISCATE"
| "chain_b_transition s B_RESTRICT   = reg_transition s RESTRICT"
\end{lstlisting}

The instance exploits two locale features simultaneously. First, the source action set parameter \isaf{actions\textsubscript{s}} is instantiated with the strict subset \isaf{escalation\_actions} of \isaf{reg\_action}, scoping structural preservation to the escalation actions only. Second, the source and target action types (\isaf{'a} versus \isaf{'b}) carry different datatypes, exercising the locale's heterogeneous action signature. The state map is the identity, since both chains share the state space.

\begin{lstlisting}
interpretation escalation_preservation:
  state_preservation
    reg_states escalation_actions reg_transition reg_terminal
    reg_states chain_b_actions chain_b_transition reg_terminal
    "id :: reg_state ⇒ reg_state" escalation_action_map
\end{lstlisting}

The naturality conditions hold by case analysis on the four escalation actions: by construction, the target transition on each \isaf{B\_X} constructor mirrors the source transition on the corresponding action. Sequential preservation then follows as a direct corollary of \Cref{thm:sequential}. Beyond serving as a witness, this instance isolates \isaf{actions\textsubscript{s}} as the formal handle for partial action vocabularies, a concern that recurs in multi-jurisdiction coordination and escalation-only consensus systems.

\subsection{Layer-Crossing Instance: On-Chain/DAML Bridge}
\label{sec:bridge-instance}

The second instance witnesses \isaf{symmetric\_state\_preservation} by modeling the bidirectional binding between two representations of the same regulatory state: the on-chain \isaf{reg\_state} enum and an off-chain DAML structured permission record.

\needspace{8\baselineskip}
\begin{lstlisting}
datatype daml_status_tag =
  D_Active | D_Frozen | D_Seized | D_Confiscated | D_Restricted

record daml_perm =
  status_tag        :: daml_status_tag
  seized_by         :: "nat option"
  restriction_scope :: "nat option"

definition valid_daml_perm :: "daml_perm ⇒ bool" where
  "valid_daml_perm p ⟷
     (status_tag p = D_Seized ⟷ seized_by p ≠ None) ∧
     (status_tag p = D_Restricted ⟷ restriction_scope p ≠ None)"
\end{lstlisting}

The auxiliary fields (\isaf{seized\_by}, \isaf{restriction\_scope}) hold the party/scope metadata required by the corresponding status, and the type-level invariant \isaf{valid\_daml\_perm} ties each auxiliary field to its status tag (non-\isaf{None} exactly when the status semantically requires it). This captures DAML's \emph{need-to-know} model, where entitlement metadata is bound to the regulatory status it implies.

The two layers process the same seven actions, so the action map is the identity in both directions; all non-trivial content lives in the layer-crossing state mapping (\Cref{fig:daml-bridge}). We take \isaf{daml\_states} to be the image \isaf{range reg\_to\_daml} and define the two maps, then register both directions as \isaf{state\_preservation} interpretations (\isaf{forward\_layer\_preservation} and \isaf{backward\_layer\_preservation}) and combine them into the symmetric bridge:

\begin{lstlisting}
interpretation onchain_daml_bridge:
  symmetric_state_preservation
    reg_states reg_actions reg_transition reg_terminal
    daml_states reg_actions daml_transition daml_terminal
    reg_to_daml id
    daml_to_reg id
\end{lstlisting}

\begin{figure}[H]
\centering
\begin{tikzpicture}[
  enum/.style={rectangle, rounded corners=2pt, draw=black, fill=gray!5, minimum width=2.6cm, minimum height=0.62cm, font=\small\sffamily},
  rec/.style={rectangle, rounded corners=2pt, draw=black, fill=gray!5, minimum width=4.6cm, minimum height=0.62cm, font=\footnotesize\ttfamily, align=left, inner sep=4pt},
  bij/.style={{Stealth[length=4pt]}-{Stealth[length=4pt]}, thick},
  meta/.style={font=\footnotesize\itshape, color=black!70},
  row/.style={node distance=0.34cm}
]
  \node[enum] (a) {ACTIVE};
  \node[enum, below=0.34cm of a] (f) {FROZEN};
  \node[enum, below=0.34cm of f] (se) {SEIZED};
  \node[enum, below=0.34cm of se] (co) {CONFISCATED};
  \node[enum, below=0.34cm of co] (re) {RESTRICTED};

  \node[rec, right=3.4cm of a]  (da) {D\_Active};
  \node[rec, right=3.4cm of f]  (df) {D\_Frozen};
  \node[rec, right=3.4cm of se] (dse){D\_Seized, seized\_by = Some p};
  \node[rec, right=3.4cm of co] (dco){D\_Confiscated};
  \node[rec, right=3.4cm of re] (dre){D\_Restricted, restriction\_scope = Some q};

  \draw[bij] (a)  -- (da);
  \draw[bij] (f)  -- (df);
  \draw[bij] (se) -- (dse);
  \draw[bij] (co) -- (dco);
  \draw[bij] (re) -- (dre);

  \node[font=\small\itshape, above=0.5cm of a]  {\isaf{reg\_state} (on-chain)};
  \node[font=\small\itshape, above=0.5cm of da] {\isaf{daml\_perm} (off-chain)};

  \node[meta] at ($(a.east)!0.5!(da.west) + (0,0.22)$) {\isaf{reg\_to\_daml} $\rightarrow$};
  \node[meta] at ($(re.east)!0.5!(dre.west) + (0,-0.24)$) {$\leftarrow$ \isaf{daml\_to\_reg}};

  \node[meta, right=0.15cm of dse] {aux};
  \node[meta, right=0.15cm of dre] {aux};
\end{tikzpicture}
\caption{The on-chain/DAML bridge as a bijection. Each \isaf{reg\_state} constructor corresponds to exactly one well-formed \isaf{daml\_perm} record (double-headed arrows), with \isaf{reg\_to\_daml} and \isaf{daml\_to\_reg} as mutual inverses on \isaf{range reg\_to\_daml}. Only \isaf{SEIZED} and \isaf{RESTRICTED} carry auxiliary metadata (party / scope), tied to the status tag by \isaf{valid\_daml\_perm}. The bijection is the content of the roundtrip guarantees, from which injectivity of \isaf{reg\_to\_daml} (no information loss) follows.}
\label{fig:daml-bridge}
\end{figure}

The roundtrip assumptions reduce to two bijection conditions, both discharged by case analysis (on \isaf{reg\_state} for one direction, on the status tag plus \isaf{valid\_daml\_perm} for the other). As a corollary, \isaf{reg\_to\_daml} is injective on \isaf{reg\_states}:

\begin{lstlisting}
corollary reg_to_daml_injective_on_states:
  "⟦ s1 ∈ reg_states; s2 ∈ reg_states;
     reg_to_daml s1 = reg_to_daml s2 ⟧ ⟹ s1 = s2"
  using onchain_daml_bridge.state_map_injective .
\end{lstlisting}

\noindent which is the formal expression of ``no information loss in the layer-crossing representation.'' That the well-formedness guard is load-bearing rather than decorative is itself machine-checked: a four-part regression witness (\isaf{guard\_is\_load\_bearing}) exhibits a record violating \isaf{valid\_daml\_perm} on which the two layers disagree about \textsc{Release}, so dropping the guard breaks naturality. Together with the state-machine and multi-domain instantiations, the safety theories contribute five concrete locale interpretations (one \isaf{state\_machine}, three \isaf{state\_preservation}, one \isaf{symmetric\_state\_preservation}) plus the parametric \isaf{multi\_domain\_preservation} instantiation. With the two D-quencer liveness interpretations of \Cref{sec:dquencer-instance} (\isaf{dq\_priority} and \isaf{dq\_fair}), the four base theories contain seven concrete locale interpretations, covering every generic locale of the safety and liveness substrate; the compositional, authenticated, and external theories (\Cref{sec:combined,sec:ads,sec:genericity}) register eleven further interpretation commands, for eighteen across the development, and every generic locale is inhabited by at least one of them.

\subsection{Comparison with the Merkle Functor Pattern}
\label{sec:merkle-comparison}

\begin{table}[H]
\centering
\caption{Structural comparison with Lochbihler's Merkle Functor~\cite{lochbihler2020}.}
\label{tab:merkle-comparison}
\small
\begin{tabular}{@{}l l l@{}}
\toprule
\textbf{Aspect} & \textbf{Merkle Functor} & \textbf{This work} \\
\midrule
Scope & Single domain (Canton) & Cross-domain, $N$ domains \\
Direction & Unidirectional & Bidirectional (roundtrip) \\
Domain count & Fixed & Arbitrary finite set \\
Isolation & N/A & \isaf{sync\_isolation} \\
Byzantine & Not included & Property~2 ($f<n/3$) \\
Locales & ADS\_Functor (1) & 10 generic locales \\
Reuse pattern & \isa{interpretation} & Same \\
\bottomrule
\end{tabular}
\end{table}

The Merkle Functor ensures data structure integrity (authentication, inclusion proofs), while the cross-domain state preservation framework presented here ensures structural preservation of state transitions. The two approaches are complementary, and we develop the composition in \Cref{sec:ads}, lifting the authenticated layer to the global-state level.

\paragraph{A note on naming.} This section establishes the foundational locales and their concrete instances. \Cref{sec:functor} promotes them to a \emph{functor} by proving the category laws (identity, composition, associativity) as theorems, so we use the term functor advisedly rather than as an analogy. The methodological lineage from Lochbihler and Mari\'{c}'s Merkle Functor is one of design philosophy (locale-based reusable abstractions over functor-shaped state machines): the safety substrate developed here is independent of the \isaf{ADS\_Functor} entry, while the authenticated layer of \Cref{sec:ads} builds on it explicitly.

\section{The Cross-Domain State Preservation Functor}
\label{sec:functor}

\Cref{sec:safety} defined structure-preserving synchronization maps between state machines. We now show that these maps are the morphisms of a category on which the construction is a \emph{functor} in a precise sense: the category laws hold as \emph{theorems}, not as an analogy. Regarding state machines as objects and the preservation maps of \isaf{State\_Preservation.thy} as morphisms, \isaf{Functor\_Laws.thy} proves identity preservation, composition preservation, and associativity.

\subsection{Category Laws on Preservation Morphisms}
\label{sec:category-laws}

\paragraph{Identity.} The identity pair $(\mathit{id}, \mathit{id})$ is a preservation morphism from any state machine to itself: a functor sends identities to identities.

\begin{lstlisting}
theorem preservation_id:
  assumes "state_machine states actions transition terminal"
  shows "state_preservation states actions transition terminal
                            states actions transition terminal id id"
\end{lstlisting}

\paragraph{Composition.} Preservation morphisms compose. Given $f : A \to B$ and $g : B \to C$, each a \isaf{state\_preservation} with its own state map and action map, the componentwise composite $(g \circ f,\ g' \circ f')$ is a preservation morphism $A \to C$.

\begin{lstlisting}
theorem preservation_compose:
  assumes "state_preservation S_a A_a T_a F_a S_b A_b T_b F_b f f'"
    and "state_preservation S_b A_b T_b F_b S_c A_c T_c F_c g g'"
  shows "state_preservation S_a A_a T_a F_a S_c A_c T_c F_c
           (g ∘ f) (g' ∘ f')"
\end{lstlisting}

The proof stacks the naturality square of the second morphism on that of the first. This is the cross-domain analogue of the ``closed under composition'' structure of Lochbihler and Mari\'{c}'s \isaf{ADS\_Functor}, whose functor is single-domain.

\paragraph{Associativity.} Composition is associative, on both the state and the action component:

\needspace{11\baselineskip}
\begin{lstlisting}
theorem preservation_assoc:
  assumes "state_preservation S_a A_a T_a F_a S_b A_b T_b F_b f f'"
    and "state_preservation S_b A_b T_b F_b S_c A_c T_c F_c g g'"
    and "state_preservation S_c A_c T_c F_c S_d A_d T_d F_d k k'"
  shows "state_preservation S_a A_a T_a F_a S_d A_d T_d F_d
           ((k ∘ g) ∘ f) ((k' ∘ g') ∘ f')
       ∧ state_preservation S_a A_a T_a F_a S_d A_d T_d F_d
           (k ∘ (g ∘ f)) (k' ∘ (g' ∘ f'))
       ∧ ((k ∘ g) ∘ f = k ∘ (g ∘ f))
       ∧ ((k' ∘ g') ∘ f' = k' ∘ (g' ∘ f'))"
\end{lstlisting}

The statement has two halves. The equational half is associativity of function composition, componentwise. The other half is the part that makes the equation a statement about the category rather than about functions: both bracketings are themselves preservation morphisms from the first object to the last, so the two sides of the equation are morphisms of the same hom-set and the equation is read there. \Cref{fig:category} depicts the three laws.

\begin{figure}[t]
\centering
\begin{tikzpicture}[
  obj/.style={rectangle, rounded corners=2pt, draw=black, fill=blue!5, minimum width=2.1cm, minimum height=0.9cm, font=\small\sffamily, align=center},
  mor/.style={-{Stealth[length=5pt]}, thick},
  node distance=3cm
]
  \node[obj] (A) {$A$};
  \node[obj, right=of A] (B) {$B$};
  \node[obj, right=of B] (C) {$C$};
  \draw[mor] (A) -- node[above, font=\footnotesize] {$f$} (B);
  \draw[mor] (B) -- node[above, font=\footnotesize] {$g$} (C);
  \draw[mor, bend right=32] (A) to node[below, font=\footnotesize] {$g \circ f$} (C);
  \draw[mor, out=125, in=55, looseness=6] (A) to node[above, font=\footnotesize] {$\mathit{id}$} (A);
\end{tikzpicture}
\caption{The category of state machines and preservation morphisms. Objects $A, B, C$ are state machines (finite states and actions, deterministic transition, terminal set); each morphism is a \isaf{state\_preservation} pair (a state map and an action map satisfying naturality). Identities exist (\isaf{preservation\_id}), morphisms compose (\isaf{preservation\_compose}), and composition is associative (\isaf{preservation\_assoc}). The construction sending each object to its synchronized image and each morphism to its componentwise action is the cross-domain state-preservation functor.}
\label{fig:category}
\end{figure}

\paragraph{A theorem, not a slogan.} We claim no new proof technique: \isaf{preservation\_assoc} reduces to associativity of function composition, and each law is discharged by unfolding the locale predicates. That cheapness is a measurement of the locale decomposition rather than of the subject, and \Cref{sec:effort} reports where the proof effort landed instead. The value is not the difficulty of any single proof but that this structure---heterogeneous-domain synchronization organized as a functor---is, to the best of our knowledge, mechanized here for the first time across domains, where Lochbihler and Mari\'{c}'s \isaf{ADS\_Functor} is single-domain. It is also the organizing foundation for what follows: the degree tower of \Cref{sec:tower} is a tower of \emph{these} functors, the convergence of \Cref{sec:combined} runs over \emph{this} category's states, and the authenticated coupling of \Cref{sec:ads} lifts an authenticated data structure onto \emph{this} functor. The functor is the spine, not a restatement.

\subsection{Transition Systems as Functors on Action Words}
\label{sec:action-functor}

The functorial reading is carried at a second place. Each transition system is itself a functor from the one-object category of \emph{action words} (the free monoid of action sequences) into the Kleisli category of the option monad on states. Sequential application is defined by

\begin{lstlisting}
fun apply_actions :: "'s ⇒ 'a list ⇒ 's option" where
  "apply_actions s [] = Some s"
| "apply_actions s (a # as) =
     (case transition s a of None ⇒ None
      | Some s' ⇒ apply_actions s' as)"
\end{lstlisting}

The identity word acts as the identity (\isaf{apply\_actions~s~[]~=~Some~s}), and a concatenation of words is the Kleisli composition of the two segment maps:

\begin{lstlisting}
lemma apply_actions_append:
  "apply_actions s (as @ bs) =
     (case apply_actions s as of None ⇒ None
      | Some s' ⇒ apply_actions s' bs)"
\end{lstlisting}

These are the two functor laws on action words. A preservation morphism carries this functor structure to the word level: if a source word is enabled, the mapped word is enabled at the target with the mapped final state (\isaf{sequential\_preservation}, \Cref{thm:sequential}). The two functorial readings---morphisms between transition systems, and each transition system as a functor on its action words---are the categorical skeleton on which \Cref{sec:tower} builds a tower of degree-indexed functors.

\section{Liveness under Byzantine Faults}
\label{sec:liveness}

\subsection{Reusable Locale Architecture}
\label{sec:liveness-locales}

The core contribution of Property~2 is a liveness verification infrastructure comprising two generic locales, defined in \isaf{Priority\_Resolution.thy} (354~lines). These locales are domain-independent.

\begin{figure}[t]
\centering
\begin{tikzpicture}[
  generic/.style={rectangle, rounded corners=2pt, draw=black, fill=blue!5, minimum width=3.8cm, minimum height=1.2cm, font=\small\sffamily, align=center},
  instance/.style={rectangle, rounded corners=2pt, draw=black, fill=orange!8, minimum width=5cm, minimum height=1.2cm, font=\small\sffamily, align=center},
  arr/.style={-{Stealth[length=5pt]}, thick},
  iarr/.style={-{Stealth[length=5pt]}, thick, dashed, color=orange!70!black},
  node distance=1.2cm and 0.5cm
]
  \node[generic] (ps) {\textbf{priority\_system}\\{\footnotesize deterministic selection}};
  \node[generic, right=2.6cm of ps] (fl) {\textbf{fair\_leader\_system}\\{\footnotesize starvation freedom}};

  \coordinate (mid) at ($(ps)!0.5!(fl)$);
  \node[instance, below=2.2cm of mid] (dq) {\textbf{DQuencer\_Instance}\\{\footnotesize regulatory domain instantiation}\\{\footnotesize + conditional\_safety\_preservation}};

  \node[generic, left=1.2cm of dq, fill=green!8] (ri) {\textbf{Regulatory\_Instance}\\{\footnotesize Property 1}};

  \draw[iarr] (ps) -- (dq);
  \draw[iarr] (fl) -- (dq);
  \draw[arr] (ri) -- (dq);

  \node[above=0.2cm of ps, font=\footnotesize\itshape, color=black!65] {Priority\_Resolution.thy};
  \node[below=0.2cm of dq, font=\footnotesize\itshape, color=black!65] {DQuencer\_Instance.thy};
\end{tikzpicture}
\caption{Locale architecture for liveness properties (Property~2). Two generic locales (blue), \isaf{priority\_system} and \isaf{fair\_leader\_system}, are defined independently of the regulatory domain in \isaf{Priority\_Resolution.thy}. \isaf{DQuencer\_Instance} (orange) instantiates both and imports \isaf{Regulatory\_Instance} (green) from Property~1; its safety-side leaf corollary is \isaf{conditional\_safety\_preservation}. The \isaf{priority\_system} instance is registered on the \isaf{priority\_key} carrier (identity projection), with the underlying messages recovered inside a well-formedness sublocale (\Cref{sec:dquencer-instance}), leaving the generic \isaf{dquencer\_system} locale unmodified.}
\label{fig:liveness-architecture}
\end{figure}

\subsubsection{priority\_system.}
Guarantees deterministic selection from a finite message set via a total order with injective priorities.

\begin{lstlisting}
locale priority_system =
  fixes priority :: "'m ⇒ 'k::linorder"
  assumes priority_injective:
    "⟦ priority m1 = priority m2 ⟧ ⟹ m1 = m2"
\end{lstlisting}

The injectivity assumption models tiebreaking: in practice, a composite key (authority level, timestamp, severity, node ID) ensures distinct messages have distinct priorities.

\begin{theorem}[Deterministic Selection]
\label{thm:deterministic}
A non-empty finite set with injective priorities has a unique maximum element.
\end{theorem}

\begin{lstlisting}
lemma highest_priority_exists:
  assumes "finite S" and "S ≠ {}"
  shows "∃!m. m ∈ S ∧
         (∀m' ∈ S. priority m' ≤ priority m)"
\end{lstlisting}

The proof combines the existence of the maximum in the finite image set (via \isaf{Max\_in}) with the uniqueness from injectivity. Deterministic selection (\isaf{select\_highest\_deterministic}) follows directly.

\subsubsection{fair\_leader\_system.}
Guarantees starvation freedom under periodic honest leader scheduling.

\needspace{16\baselineskip}
\begin{lstlisting}
locale fair_leader_system =
  fixes leader_at :: "nat ⇒ 'n"
    and is_honest :: "'n ⇒ bool"
    and pending :: "nat ⇒ nat"
    and fairness_bound :: nat
  assumes fair_leader:
      "∀epoch. ∃e. epoch ≤ e
        ∧ e < epoch + fairness_bound
        ∧ is_honest (leader_at e)"
    and honest_progress:
      "⟦ is_honest (leader_at e);
         pending e > 0 ⟧
       ⟹ pending (Suc e) < pending e"
    and non_honest_bounded:
      "pending (Suc e) ≤ pending e"
\end{lstlisting}

Positivity of the fairness bound is derived rather than assumed: the window starting at any epoch must be non-empty to contain the honest slot that \isaf{fair\_leader} promises, so \isaf{fairness\_bound\_positive} is a lemma of the locale, and the three assumptions shown are exactly the locale's assumptions. The \isaf{fair\_leader} assumption is a deterministic abstraction of the probabilistic guarantees of VRF-based leader election. Under $f < n/3$ Byzantine faults, the probability that a Byzantine leader is elected $k$ consecutive times is $(f/n)^k < (1/3)^k$, which decreases exponentially. The fairness bound $k$ captures this as a deterministic upper bound. This is a standard assume-guarantee reasoning pattern.

\begin{theorem}[Starvation Bound]
\label{thm:starvation-bound}
If there are pending requests, at least one is processed within \isaf{fairness\_bound} epochs.
\end{theorem}

\begin{lstlisting}
theorem starvation_bound:
  assumes "pending epoch > 0"
  shows "∃e. epoch ≤ e
            ∧ e < epoch + fairness_bound
            ∧ pending (Suc e) < pending e"
\end{lstlisting}

The proof splits on whether the honest leader epoch~$h$ (guaranteed by \isaf{fair\_leader}) has positive pending count. If $\isaf{pending}~h > 0$, \isaf{honest\_progress} gives a direct decrease. If $\isaf{pending}~h = 0$, then the pending count dropped from positive (at \isaf{epoch}) to zero (at~$h$), and an induction on the interval locates the strict decrease step.

\begin{theorem}[Eventual Completion]
\label{thm:eventual}
All pending requests are eventually processed.
\end{theorem}

\begin{lstlisting}
theorem eventual_completion:
  shows "∃e_final. pending e_final = 0"
\end{lstlisting}

The proof uses well-founded induction on the pending count (a natural number). Repeated application of \isaf{starvation\_bound} shows that the pending count strictly decreases every \isaf{fairness\_bound} epochs; by the well-foundedness of natural numbers, it reaches zero.

\subsection{Regulatory Instance under Byzantine Faults}
\label{sec:dquencer-instance}

\isaf{DQuencer\_Instance.thy} (651~lines) instantiates the generic locales with the regulatory domain.

\paragraph{Authority level.} Regulatory authority hierarchy reflecting jurisdictional priority:

\begin{lstlisting}
datatype authority_level =
  Regional | National | International
\end{lstlisting}

\paragraph{Priority key.} A 4-tuple of natural numbers using Isabelle's built-in lexicographic product order:

\begin{lstlisting}
type_synonym priority_key = "nat × nat × nat × nat"
\end{lstlisting}

Components (highest to lowest significance): (1)~\isaf{authority\_rank}, (2)~inverted timestamp (earlier = higher), (3)~\isaf{action\_severity}, (4)~inverted node ID (deterministic tiebreaker). Injectivity is proven (\isaf{priority\_key\_injectivity}).

\paragraph{BFT system locale.} The consensus system itself is a locale carrying the node roster, the fault bound, and the scheduling and precision parameters:

\needspace{18\baselineskip}
\begin{lstlisting}
locale dquencer_system =
  fixes nodes :: "node_info set"
    and f_max :: nat
    and fairness_bound :: nat
    and max_time :: nat
    and max_node :: nat
  assumes finite_nodes: "finite nodes"
    and bft_threshold:
      "card nodes ≥ 3 * f_max + 1"
    and byzantine_bound:
      "card (byzantine_nodes nodes) ≤ f_max"
    and fairness_positive: "fairness_bound > 0"
\end{lstlisting}

Honest majority is derived within this locale:

\begin{lstlisting}
lemma honest_majority:
  "card (honest_nodes nodes) > 2 * f_max"
\end{lstlisting}

\paragraph{Locale instantiations.} Both generic liveness locales are instantiated for the regulatory consensus system. The \isaf{fair\_leader\_system} instantiation (\isaf{dq\_fair}) is performed in a separate \isaf{dquencer\_liveness} locale that adds the leader schedule and pending count as parameters, with the honest-leader predicate mapped to \isaf{ni\_behavior}~$=$~\isaf{Honest}. The \isaf{priority\_system} instantiation is treated next.

\paragraph{Priority instance with well-formedness bounds.} The \isaf{priority\_system} locale assumes priority injectivity \emph{unconditionally} on its carrier. For D-quencer messages this is delicate: injectivity of \isaf{make\_priority\_key} holds only under two well-formedness bounds, $\isaf{msg\_timestamp} \le \isaf{max\_time}$ and $\isaf{dqm\_source\_node} \le \isaf{max\_node}$, which reflect a fixed-precision clock and a fixed node roster but are not properties of the \isaf{dq\_message} type as such. Rather than impose these bounds at the type level, we instantiate \isaf{priority\_system} directly on the \isaf{priority\_key} type with the identity projection, so injectivity holds by reflexivity:

\begin{lstlisting}
interpretation dq_priority:
  priority_system "id :: priority_key ⇒ priority_key"
  by unfold_locales auto
\end{lstlisting}

The connection back to \isaf{dq\_message} is then made inside two sublocales that carry the bounds explicitly. \isaf{dquencer\_priority\_context} extends \isaf{dquencer\_system} with a well-formed message set and defines the induced priority-key set; \isaf{dquencer\_priority\_concrete} adds priority-key distinctness (the natural BFT precondition on authority/timestamp/severity/node tuples). Under that assumption the key uniquely identifies its message, captured by a recovery function with a unicity lemma:

\begin{lstlisting}
definition recover_msg ::
  "priority_key ⇒ dq_message" where
  "recover_msg k =
     (THE m. m ∈ msg_set
        ∧ make_priority_key max_time max_node m = k)"
\end{lstlisting}

Running the generic \isaf{select\_highest} on the priority-key set and then applying \isaf{recover\_msg} yields a deterministic choice of the highest-priority well-formed message, exposed as the corollaries \texttt{dq\_select\_\allowbreak highest\_\allowbreak deterministic} and \texttt{dq\_select\_\allowbreak highest\_\allowbreak message}. Crucially, the \isaf{dquencer\_system} locale itself is not modified: the bounds live in the sublocales, keeping the generic priority abstraction intact.

\subsection{Assume-Guarantee Reasoning for Starvation Freedom}
\label{sec:assume-guarantee}

The starvation freedom guarantee follows an assume-guarantee structure:

\begin{itemize}[leftmargin=*]
\item \textbf{Assumption:} within any \isaf{fairness\_bound} consecutive epochs, at least one epoch has an honest leader (\isaf{fair\_leader}).
\item \textbf{Guarantee:} under this assumption, all pending requests are processed within bounded time (deterministic proof).
\item \textbf{Justification of the assumption:} the probability of $k$ consecutive Byzantine leaders is $(f/n)^k$, which decreases exponentially. For $f < n/3$ and $k = 10$: $(1/3)^{10} \approx 1.7 \times 10^{-5}$. VRF literature provides the probabilistic backing (outside the proof scope).
\end{itemize}

\subsection{The Byzantine Threshold Is Load-Bearing}
\label{sec:inhabitability}

A liveness locale that no concrete system satisfies would be vacuous. We therefore show that the Byzantine threshold $n \geq 3f+1$ is \emph{load-bearing}: it drives a chain of lemmas that makes the assume-guarantee liveness locale inhabitable, and we exhibit a concrete non-degenerate instance witnessing it.

\begin{figure}[t]
\centering
\begin{tikzpicture}[
  lem/.style={rectangle, rounded corners=2pt, draw=black, fill=blue!5, minimum width=3.4cm, minimum height=0.85cm, font=\footnotesize\sffamily, align=center},
  goal/.style={rectangle, rounded corners=2pt, draw=black, fill=green!10, minimum width=3.4cm, minimum height=0.9cm, font=\footnotesize\sffamily, align=center},
  wit/.style={rectangle, rounded corners=2pt, draw=black, fill=orange!8, minimum width=3.6cm, minimum height=1.0cm, font=\footnotesize\sffamily, align=center},
  arr/.style={-{Stealth[length=5pt]}, thick},
  warr/.style={-{Stealth[length=5pt]}, thick, dashed, color=orange!70!black},
  node distance=0.62cm
]
  \node[lem] (bft) {\isaf{bft\_threshold}\\{\footnotesize $n \geq 3f+1$}};
  \node[lem, below=of bft] (maj) {\isaf{honest\_majority}\\{\footnotesize honest ${}> 2f$}};
  \node[lem, below=of maj] (ne) {\isaf{honest\_nonempty}};
  \node[lem, below=of ne] (fs) {\isaf{fair\_schedule\_exists}};
  \node[goal, below=of fs] (inh) {\isaf{liveness\_inhabitable}\\{\footnotesize locale is satisfiable}};

  \draw[arr] (bft) -- (maj);
  \draw[arr] (maj) -- (ne);
  \draw[arr] (ne) -- (fs);
  \draw[arr] (fs) -- (inh);

  \node[wit, right=2.6cm of ne] (q) {\isaf{bft\_quorum} witness\\{\footnotesize 4 nodes, 1 Byzantine}\\{\footnotesize tight $4 \geq 3\cdot 1+1$}};
  \draw[warr] (q) -- (inh);
\end{tikzpicture}
\caption{The Byzantine threshold is load-bearing. The chain from \isaf{bft\_threshold} to \isaf{liveness\_inhabitable} makes the assume-guarantee liveness locale satisfiable; deleting any link breaks the proof. The \isaf{bft\_quorum} interpretation (right) is a concrete non-degenerate witness: four nodes with one Byzantine node at the tight bound, exercising a genuine (non-vacuous) progress obligation. The headline bounded-convergence theorem (\Cref{sec:combined}) does not route through this inhabitability witness; it is driven instead by the cross-chain inconsistency measure.}
\label{fig:threshold-loadbearing}
\end{figure}

From \isaf{bft\_threshold} ($n \geq 3f+1$), honest majority (more than $2f$ honest nodes) follows; hence the honest set is non-empty, hence a constant-honest fair schedule exists. This schedule instantiates the assume-guarantee liveness locale, proving it satisfiable (\isaf{liveness\_inhabitable}). Each link is load-bearing: deleting any of \isaf{bft\_threshold}, \isaf{honest\_majority}, \isaf{honest\_nonempty}, or \isaf{fair\_schedule\_exists} breaks the proof. Satisfiability is further witnessed concretely by \isaf{bft\_quorum}, a four-node instance with one Byzantine node at the tight bound $4 \geq 3\cdot 1+1$, exercising a non-degenerate progress obligation rather than a vacuous one. This inhabitability result is a satisfiability witness for the locale; the headline convergence theorem of \Cref{sec:combined} does not route through it.

\section{Combined Safety and Liveness: Composition and Convergence}
\label{sec:combined}

The combination of Properties~1 and~2 is analyzed at two levels. A leaf corollary turns safety into an existence statement, discharging the unlocked side condition from global validity (\Cref{sec:combined-theorem}); it does not fuse liveness into safety. The genuine fusion is a bounded-convergence result of the compositional layer, which frees the safety guarantee of its initial-consistency hypothesis through a well-founded measure on cross-chain inconsistency, under the fair-leader assumption (\Cref{sec:discharge-structure}).

\begin{figure}[t]
\centering
\begin{tikzpicture}[
  box/.style={rectangle, rounded corners=2pt, draw=black, minimum width=5.2cm, minimum height=1.3cm, font=\small\sffamily, align=center},
  arr/.style={-{Stealth[length=6pt]}, very thick},
  node distance=1.6cm and 0.9cm
]
  \node[box, fill=blue!8] (p1) {\textbf{Property 1 (Safety)}\\{\footnotesize \emph{if} valid initial state and honest execution}\\{\footnotesize \emph{then} cross-domain state preservation}};
  \node[box, fill=orange!8, right=of p1] (p2) {\textbf{Property 2 (Liveness)}\\{\footnotesize under $f < n/3$ and a fair-leader bound:}\\{\footnotesize determinism ${+}$ starvation freedom}};

  \node[box, fill=gray!8, below=1.7cm of p1] (leaf) {\textbf{conditional\_safety\_preservation}\\{\footnotesize leaf corollary: safety as existence,}\\{\footnotesize unlocked side condition discharged}};
  \node[box, fill=green!10, below=1.7cm of p2] (conv) {\textbf{guarded bounded convergence}\\{\footnotesize genuine fusion: from an arbitrary}\\{\footnotesize unlocked state, no initial consistency}\\{\footnotesize assumed, converges to a valid state}};

  \draw[arr, color=blue!70!black] (p1) -- node[left, font=\footnotesize\itshape] {restate} (leaf);
  \draw[arr, color=green!45!black] (p1) -- node[near end, above, sloped, font=\footnotesize\itshape] {inconsistency measure} (conv);
  \draw[arr, color=orange!70!black] (p2) -- node[right, font=\footnotesize\itshape] {fair leader} (conv);
\end{tikzpicture}
\caption{How the two properties compose. Property~1's safety guarantee is turned by the leaf corollary \isaf{conditional\_safety\_preservation} into an existence statement, with the unlocked side condition discharged from global validity rather than assumed; this is not a fusion of liveness into safety. The genuine fusion is \emph{guarded bounded convergence}: combining the safety-side step with the fair-leader liveness assumption and a well-founded measure on cross-chain inconsistency, synchronization converges to a valid state from an arbitrary unlocked initial configuration, with no assumption that cross-chain consistency holds initially. This convergence result lives in the compositional layer.}
\label{fig:discharge}
\end{figure}

\subsection{The Convergence Structure}
\label{sec:discharge-structure}

Property~1's guarantee is conditional: ``\emph{if} the synchronization coordinator executes correctly \emph{and} the initial global state is valid, \emph{then} cross-domain state is preserved.'' Two questions then arise: what happens when the initial state is \emph{not} assumed valid, and what does liveness add?

Property~2 supplies liveness in a Byzantine environment ($f < n/3$): determinism (BFT consensus with honest majority produces a deterministic result) and starvation freedom (fair leader scheduling ensures every pending request is processed). It does not by itself discharge the honest-execution premise of Property~1.

The genuine combination is a \emph{convergence} result rather than an assumption discharge. Working over the compositional layer, one drops the initial-consistency hypothesis and instead equips the global state with a well-founded measure---the number of cross-chain inconsistency pairs. Under the fair-leader assumption, each honest step strictly decreases this measure until it reaches zero, at which point the state is valid. The outcome is bounded convergence to a valid state from an arbitrary unlocked configuration, at the price of the fair-leader context; it is not an unconditional guarantee. Where Property~1 and its leaf corollary \emph{assume} a valid starting state and preserve it, convergence assumes none: it establishes that the system \emph{reaches} validity from any unlocked configuration. This self-healing guarantee, rather than invariant preservation, is the precise sense in which liveness strengthens safety.

\subsection{The Generic Convergence Locale}
\label{sec:converging-locale}

Convergence is proved generically in \isaf{Composition.thy}, then instantiated on the oracle. The theory composes a guarded safety invariant with an eventual-progress scheduler and adds a natural-number progress measure. Its obligations are spread over three layers: a guarded invariant (a guarded step preserves the invariant), bounded fairness (within a fixed \isaf{window}, some scheduled event discharges), and strict measure decrease on every discharging step taken from a non-invariant state:

\begin{lstlisting}
locale converging_composition = ... +
  fixes progress_measure :: "'s ⇒ nat"
  assumes discharge_progresses:
    "⟦ s ∈ carrier; ¬ inv s; discharges ev ⟧
     ⟹ ∃opn s'. realize ev s = Some opn
                 ∧ step s opn = Some s'
                 ∧ progress_measure s' < progress_measure s"
\end{lstlisting}

Two properties one might assume are instead \emph{derived}: that the window is positive (from bounded occurrence), and that the measure's zero set lies inside the invariant (otherwise a discharging step at measure zero would force a negative target). Well-foundedness is supplied by the natural-number order. The headline is convergence from an \emph{arbitrary} carrier state, within a bound set by the measure and the window:

\begin{lstlisting}
definition convergence_bound :: "'s ⇒ nat" where
  "convergence_bound s = progress_measure s * window"

theorem bounded_convergence_from_arbitrary:
  assumes "s ∈ carrier"
  shows "∃t ≤ convergence_bound s. ∃s'.
           evolves_to s t s' ∧ inv s'"
\end{lstlisting}

The theorem assumes nothing about the invariant holding initially: it converges to it. The oracle instance below arises by mapping the carrier to the finite-domain, unlocked global states, the invariant to \isaf{consistent\_state}, the discharger to honest-leader epochs, the window to \isaf{fairness\_bound}, and---the semantic core---the progress measure to \isaf{inconsistency\_pairs}, the number of cross-chain inconsistency pairs. Nothing in the composition is free: the price of dropping the initial-consistency hypothesis is the fair-leader assumption that the scheduler supplies.

\subsection{The Leaf Corollary and the Convergence Theorem}
\label{sec:combined-theorem}

The safety-side leaf corollary upgrades \isaf{valid\_state\_preservation} from an implication about a successful synchronization to an existence statement: from a valid global state, on an asset whose regulatory transition is enabled, the synchronization \emph{does} succeed and the result is again valid. No unlocked side condition is assumed. It is discharged instead: global validity contains \isaf{no\_locked\_without\_reason}, from which $\lnot\,\isaf{is\_locked}~\isaf{gs}~\isaf{aid}$ follows, and \isaf{lock\_acquire\_success} then supplies the lock. The proof uses only the safety side; it invokes none of the liveness interpretations and makes no Byzantine, determinism, or starvation claim of its own.

\begin{lstlisting}
theorem conditional_safety_preservation:
  assumes valid: "valid_state gs"
    and current:
      "get_reg_state gs source aid = Some s"
    and trans:
      "reg_transition s action = Some s'"
  shows "∃gs'. sync source action aid gs
               = Some gs' ∧ valid_state gs'"
\end{lstlisting}

The genuine safety--liveness fusion is a separate theorem of the compositional layer. It drops the \isaf{valid\_state} hypothesis assumed above and converges to a valid state from an arbitrary unlocked configuration, within a bound set by the initial inconsistency count and the fairness bound:

\begin{lstlisting}
theorem oraclizer_guarded_bounded_convergence:
  assumes "finite_domain gs_0"
    and "no_locked_without_reason gs_0"
  shows "∃t ≤ oss.convergence_bound gs_0. ∃gs_t.
           oss.evolves_to gs_0 t gs_t ∧ valid_state gs_t"
\end{lstlisting}

Here the well-founded measure is \isaf{inconsistency\_pairs} (the number of cross-chain inconsistency pairs), which each honest step strictly decreases under the fair-leader assumption until the state is valid.

\subsection{Terminal-Faithful Recovery}
\label{sec:terminal-faithful}

Convergence says that the system reaches \emph{a} valid state; it does not by itself pin down \emph{which} one. Two mechanically checked counterexamples showed that converged-and-valid alone underdetermines the direction of reconciliation: a rogue recovery could reach validity by confiscating every disagreeing asset outright, or by overwriting a recorded confiscation with a non-terminal value and resurrecting the asset everywhere. The recovery path is therefore constrained on the terminal axis. The predicate \isaf{safe\_recovery} requires every reconciling synchronization to strictly decrease the inconsistency measure and to issue \textsc{Confiscate} exactly when a confiscation is already recorded on some chain (\isaf{terminal\_present}). Two results then pin the recovery layer to the only behaviours the regulatory semantics admits for \textsc{Confiscated}:

\begin{lstlisting}
theorem recovery_no_fresh_terminal:
  assumes "gs_0 ∈ oss_carrier"
    and "¬ terminal_present gs_0 aid"
    and "oss.evolves_to gs_0 t gs_t"
  shows "¬ terminal_present gs_t aid"
\end{lstlisting}

\noindent an asset that starts with no recorded confiscation never acquires one along any evolution, however the run is scheduled; and \isaf{recovery\_terminal\_completion}: on a terminal-bearing asset a safe recovery can only be the completing confiscation, so an existing confiscation is propagated, never overwritten. Terminal facts, the regulatory finality of \Cref{sec:locale-hierarchy}, are invariant along the healing path: convergence restores consistency without rewriting enforcement history. These theorems bound the recovery path from below; a full direction-of-reconciliation semantics, designating which record is authoritative when histories disagree, is left to subsequent work.

\subsection{What the Combination Establishes}
\label{sec:what-combined}

\begin{enumerate}[leftmargin=*]
\item \textbf{Safety} (Property~1): after synchronization, all connected chains agree on the regulatory state. The global validity invariant is preserved.
\item \textbf{Determinism} (Property~2): conflicting regulatory actions are resolved by a total order on priority keys; the consensus output is unique.
\item \textbf{Starvation freedom} (Property~2): under fair leader scheduling, every pending regulatory request is processed within a bounded number of epochs.
\item \textbf{Bounded convergence} (compositional layer): from an arbitrary unlocked state, with no assumption that cross-chain consistency holds initially, synchronization reaches a valid state within a bounded number of steps under the fair-leader assumption, along a terminal-faithful recovery path (\Cref{sec:terminal-faithful}).
\end{enumerate}

Deadlock is a concurrency phenomenon; under the atomic synchronization model there is no concurrent lock contention, so the development states no deadlock-freedom theorem. Forced lock release under contention is deferred to the preemptive-lock property (future work).

\subsection{Implementation Parameter Design Guide}
\label{sec:impl-guide}

Parameter constraints derived from the proofs translate directly into implementation design rationale:

\begin{table}[ht]
\centering
\caption{Implementation parameters derived from proofs.}
\label{tab:impl-params}
\small
\begin{tabular}{@{}l l l@{}}
\toprule
\textbf{Parameter} & \textbf{Formal Constraint} & \textbf{Implementation Consideration} \\
\midrule
BFT threshold & $n \geq 3f + 1$ & 20 nodes $\Rightarrow$ max 6 Byzantine \\
Fairness bound & $k$ honest within $k$ & VRF parameters, leader rotation \\
Priority order & Injective 4-tuple & authority $\times$ time $\times$ severity $\times$ nodeID \\
\bottomrule
\end{tabular}
\end{table}

\section{The Synchronization-Degree Functor Tower}
\label{sec:tower}

The product classifies each asset by the \emph{synchronization degree} it requires, from $S_0$ (static, no synchronization) to $S_3$ (atomic state binding), and the development formalizes this hierarchy as a tower of functors indexed by degree. One caveat governs the whole section: $F(k)$ formalizes the \emph{coupling breadth} of degree $k$---which chains lie within synchronization reach---while the per-degree synchronization \emph{semantics} itself (static, unidirectional observation, bidirectional coupling, atomic binding) is abstracted, not formalized. What is proved is the categorical structure of the ladder, not the operational meaning of each rung.

\subsection{Degree Functors and Their Functor Laws}
\label{sec:degree-functors}

For each degree $k$, $F(k)$ packages a carrier (\isaf{deg\_carrier~k}, the global states whose asset-bearing chains all lie within reach $\le k$) with a step function \isaf{deg\_step~k} that applies a regulatory transition at the hub and broadcasts the result to every chain within reach. Its action on a word of regulatory actions is

\begin{lstlisting}
fun deg_run :: "nat ⇒ asset_id ⇒ reg_action list
               ⇒ global_state ⇒ global_state option"
\end{lstlisting}

\noindent and it satisfies the two functor laws on action words---the identity word acts as the identity, and concatenation is Kleisli composition:

\begin{lstlisting}
lemma deg_run_Nil: "deg_run k aid [] = Some"

lemma deg_run_append:
  "deg_run k aid (v @ w) gs
     = Option.bind (deg_run k aid v gs) (deg_run k aid w)"
\end{lstlisting}

These two laws are the categorical skeleton. The semantic content---that the functor is \emph{non-degenerate} rather than a vacuous fold over an index set---is a collapse theorem: a non-empty word of degree-$k$ synchronizations on an asset held at the hub equals broadcasting the \emph{composite} regulatory state once.

\begin{lstlisting}
lemma deg_run_collapse:
  assumes "get_reg_state gs 0 aid = Some s0"
  shows "deg_run k aid (a # w) gs
         = (case reg_run s0 (a # w) of None ⇒ None
            | Some sn ⇒ Some (broadcast_le k aid sn gs))"
\end{lstlisting}

The collapse is not a free-monoid triviality. It rests on a broadcast-overwrite lemma---a second degree-$k$ broadcast of the same asset obliterates the first, because the broadcast set is exactly the chains already holding the asset, which the first broadcast does not change---and on hub tracking, so the induction genuinely uses the degree/broadcast structure. That the functor transports real regulatory content is witnessed concretely: the two-action word \textsc{Freeze} then \textsc{Confiscate} composes through the intermediate \textsc{Frozen} to a broadcast of \textsc{Confiscated}, which differs from the broadcast of \textsc{Frozen} produced by \textsc{Freeze} alone. The composition is not idempotent padding.

\subsection{Natural Transformations and the Tower}
\label{sec:nat-tower}

Adjacent degrees are connected by a forgetful map \isaf{degree\_forget}, which drops the chain at the demoted index. It is a natural transformation between the degree functors: on generators its naturality square commutes in both the enabled and the disabled case, because the hub chain is never the dropped one and so the transition's enabling condition is untouched.

\begin{lstlisting}
theorem degree_natural_transformation:
  "natural_transformation (F (Suc k)) (F k)
     (degree_forget (Suc k))"
\end{lstlisting}

The single-step square lifts, by induction on the action word, to naturality over the \emph{whole} action category, promoting the transformation from a square on generators to a genuine natural transformation between the degree functors:

\begin{lstlisting}
theorem degree_forget_natural_run:
  assumes "gs ∈ deg_carrier (Suc k)"
  shows "deg_run k aid w (degree_forget (Suc k) gs)
         = map_option (degree_forget (Suc k))
                      (deg_run (Suc k) aid w gs)"
\end{lstlisting}

Natural transformations are closed under vertical composition---the analogue of \isaf{preservation\_compose} one level up, at the level of global states---so the whole ladder $S_3 \Rightarrow S_2 \Rightarrow S_1 \Rightarrow S_0$ is natural at once (\Cref{fig:tower}):

\needspace{8\baselineskip}
\begin{lstlisting}
theorem nt_compose:
  "natural_transformation (F (Suc (Suc k))) (F k)
     (degree_forget (Suc k) ∘ degree_forget (Suc (Suc k)))"
\end{lstlisting}

Because the degree functors share one action category as common source, vertical composition is the only composition that arises for this tower; horizontal composition and whiskering do not typecheck here and are not claimed. This functor-tower layer---genuine functors with natural transformations between them---is, to our knowledge, the region that Lochbihler and Mari\'{c}'s \isaf{ADS\_Functor} does not develop: it closes functors under composition but does not build a hierarchy of functors with natural transformations between them. Demotion moreover lands inside the partial-view preorder of \Cref{sec:ads} (\isaf{degree\_forget\_refines}), so the hierarchy sits \emph{on top of} the functor laws and the authenticated layer, not beside them.

\begin{figure}[t]
\centering
\begin{tikzpicture}[
  lvl/.style={rectangle, rounded corners=2pt, draw=black, fill=blue!5, minimum width=5.4cm, minimum height=0.9cm, font=\small\sffamily, align=center},
  nt/.style={-{Stealth[length=5pt]}, very thick, color=orange!70!black},
  node distance=1.0cm
]
  \node[lvl] (f3) {$F(3)$ \; {\footnotesize $S_3$: atomic state binding}};
  \node[lvl, below=of f3] (f2) {$F(2)$ \; {\footnotesize $S_2$: bidirectional coupling}};
  \node[lvl, below=of f2] (f1) {$F(1)$ \; {\footnotesize $S_1$: unidirectional observation}};
  \node[lvl, below=of f1] (f0) {$F(0)$ \; {\footnotesize $S_0$: static, no synchronization}};
  \draw[nt] (f3) -- node[right, font=\footnotesize] {\isaf{degree\_forget}} (f2);
  \draw[nt] (f2) -- node[right, font=\footnotesize] {\isaf{degree\_forget}} (f1);
  \draw[nt] (f1) -- node[right, font=\footnotesize] {\isaf{degree\_forget}} (f0);
  \draw[dashed, gray!70] ($(f2.south west)+(-0.5,0)$) -- ($(f2.south east)+(2.6,0)$)
    node[right, font=\footnotesize\itshape, color=black!65, align=left] {causal boundary\\($k = 2$)};
\end{tikzpicture}
\caption{The synchronization-degree functor tower. Each $F(k)$ is a functor on regulatory action words formalizing the coupling breadth of degree $k$; consecutive levels are connected by the forgetful natural transformation \isaf{degree\_forget}, closed under vertical composition (\isaf{nt\_compose}), so the whole chain is natural at once. The per-degree synchronization semantics ($S_0$--$S_3$ labels) is abstracted, not formalized; only the ladder's categorical structure is proved. The causal-consistency boundary sits at $k = 2$, the $S_1/S_2$ transition.}
\label{fig:tower}
\end{figure}

\subsection{Degree Monotonicity}
\label{sec:monotonicity}

The intended safety statement is that provisioning a higher degree than an asset requires is safe, while provisioning a lower one is not. Care is needed here, twice over. First, mere validity preservation under processing is \emph{degree-free}---it holds regardless of the provisioned degree---so the lemma expressing it (\isaf{hierarchy\_monotonicity}) is recorded as an auxiliary aliasing that degree-free fact, \emph{not} as the headline. Second, the same caution applies to the valid-state form of the positive direction: on a valid state every chain already agrees with the hub, so broadcasting the hub value reconciles trivially at \emph{any} degree, and the theorem

\begin{lstlisting}
theorem over_provisioning_guarantees:
  assumes "asset_degree aid ≤ d" and "valid_state gs"
  shows "guarantees_preservation d gs aid"
\end{lstlisting}

\noindent documents the intended provisioning regime rather than exercising its degree hypothesis. The genuinely degree-sensitive, one-directional monotonicity is carried by a matched pair over \emph{arbitrary} hub-defined states. Over-provisioning drives every required chain to the single hub value, from any state on which the hub holds the asset, consistent or not:

\begin{lstlisting}
theorem over_provisioning_reconciles:
  assumes "asset_degree aid ≤ d"
    and "get_reg_state gs 0 aid = Some v"
  shows "guarantees_preservation d gs aid"
\end{lstlisting}

\noindent while under-provisioning carries no guarantee at all---some state defeats every preservation claim:

\begin{lstlisting}
theorem no_downward_safety:
  assumes "asset_degree aid > system_degree"
  shows "¬ (∀gs. guarantees_preservation system_degree gs aid)"
\end{lstlisting}

The witness for \isaf{no\_downward\_safety} places the hub at \textsc{Active} and a required chain just beyond the system's reach at \textsc{Frozen}; processing reconciles only the chains within reach, leaving the out-of-reach chain in disagreement. \isaf{over\_provisioning\_reconciles} succeeds on exactly this class of disagreeing states when the degree suffices, and \isaf{no\_downward\_safety} exhibits the failure on one of them when it does not. Both directions quantify over arbitrary states, so the contrast pair is matched and the monotonicity is genuinely one-directional; dropping the degree hypothesis from the former would be refuted by the same disagreeing witness. Here \isaf{asset\_degree} is a context parameter (every theorem holds for every assignment) and \isaf{system\_degree} is universally quantified; the statements therefore speak of an arbitrary system capability degree.

\subsection{The Causal-Consistency Boundary}
\label{sec:causal-boundary}

The degree classes $S_1$ and $S_2$ are separated by whether the asset requires causal consistency: degree $\ge 2$ does, lower degrees do not. We model the happened-before relation by the strict order on integer timestamps---a plain strict partial order (irreflexive, asymmetric, transitive), \emph{not} a distributed causal order---and prove the boundary well defined, with over-provisioning respecting it:

\begin{lstlisting}
theorem boundary_well_defined:
  "(causal_consistent_at aid d
      ⟷ (2 ≤ asset_degree aid ⟶ 2 ≤ d))
   ∧ (asset_degree aid ≤ d ⟶ causal_consistent_at aid d)
   ∧ (∀t1 t2. lamport_hb t1 t2 ⟶ ¬ lamport_hb t2 t1)
   ∧ (∀t. ¬ lamport_hb t t)
   ∧ (∀t1 t2 t3. lamport_hb t1 t2 ⟶ lamport_hb t2 t3
                 ⟶ lamport_hb t1 t3)"
\end{lstlisting}

Static promotion---reassigning an asset's degree between synchronizations---is just a different assignment parameter and is covered by the parametricity in \isaf{asset\_degree}. In-flight promotion, a reassignment crossing a live synchronization, is out of scope here and left to subsequent work on retry-queue semantics.

\section{Coupling to an Authenticated Data Structure}
\label{sec:ads}

The state-preservation functor is coupled to an authenticated data structure by building directly on Lochbihler and Mari\'{c}'s \isaf{ADS\_Functor} entry~\cite{lochbihler2020}. Whereas that entry's functor is single-domain, we lift its merge and blinding operations to the cross-domain global-state level through two glue predicates, and instantiate the coupling both on the blindable-position functor and on a recursive model of the Canton transaction tree. It is worth being exact about where that dependency bites, because ``builds on an AFP entry'' can mean anything from a session declaration to a proof step. Three levels should be distinguished. At the \emph{session} level, \isaf{ADS\_Functor} is a declared dependency of the whole development, so every theorem is checked in its presence. At the \emph{interface} level, two theories instantiate its \isaf{merkle\_interface} locale: \isaf{Functor\_Laws.thy}, through the \isaf{authenticated\_state} locale and its \isaf{oss\_authenticated} interpretation, and \isaf{Canton\_Bridge.thy}, through \isaf{oss\_blindable}, \isaf{canton\_authenticated} and \isaf{reg\_tx\_authenticated}. At the \emph{proof} level---the only one that makes the dependency mathematical rather than architectural---the interface's own laws appear as named steps inside our proofs: \isaf{authenticated\_preservation\_soundness} and \isaf{blinded\_view\_preserves\_validity} invoke \isaf{merge\_respects\_hashes}, \isaf{join} and \isaf{hash}, and \isaf{cdsp\_ads\_merge\_assoc} invokes the interface's associativity and commutativity laws. Outside \isaf{Functor\_Laws.thy} and \isaf{Canton\_Bridge.thy} the dependency is architectural only: the safety substrate of \Cref{sec:safety,sec:functor} proves its theorems without it, and \Cref{sec:tower} reaches the authenticated layer transitively, through \isaf{state\_refines}.

\subsection{State-Level Glue: Join and Refinement}
\label{sec:glue}

Two predicates read the ADS operations at the level of global states (\Cref{fig:ads}). \isaf{state\_refines}~$s_a$~$s_b$ says that $s_a$ is a partial view of $s_b$---it agrees with $s_b$ on every chain it can observe---which is the state-level reading of the Merkle blinding order:

\needspace{12\baselineskip}
\begin{lstlisting}
definition state_refines ::
  "global_state ⇒ global_state ⇒ bool" where
  "state_refines sa sb ≡
     ∀aid c. c ∈ connected_chains sa aid
       ⟶ c ∈ connected_chains sb aid
         ∧ get_reg_state sa c aid = get_reg_state sb c aid"
\end{lstlisting}

It is a preorder (\isaf{state\_refines\_refl}, \isaf{state\_refines\_trans}) and preserves consistency. Dually, \isaf{state\_join}~$s_a$~$s_b$~$s_{ab}$ says that $s_{ab}$ is the consistent combination of the partial views $s_a$ and $s_b$: it agrees with each on its own chains, the two views agree where they overlap, and---a containment clause that does real work---$s_{ab}$ carries no chain outside both views. This is the state-level reading of the ADS merge (least upper bound). The containment clause is load-bearing: a regression witness (\isaf{rogue\_join\_excluded}) rejects a candidate that smuggles in a chain neither view authenticates.

\begin{figure}[t]
\centering
\begin{tikzpicture}[
  box/.style={rectangle, rounded corners=2pt, draw=black, minimum width=3.6cm, minimum height=1.7cm, font=\small\sffamily, align=center},
  arr/.style={-{Stealth[length=6pt]}, very thick},
  node distance=3.4cm
]
  \node[box, fill=gray!8] (ads) {\textbf{ADS / Merkle}\\[2pt]{\footnotesize data $d$, hash $h$}\\{\footnotesize blinding order \isaf{bo}}\\{\footnotesize merge \isaf{m} (join)}};
  \node[box, fill=green!10, right=of ads] (gs) {\textbf{Global states}\\[2pt]{\footnotesize \isaf{valid\_state}}\\{\footnotesize refinement \isaf{state\_refines}}\\{\footnotesize combination \isaf{state\_join}}};
  \draw[arr] (ads) -- node[above, font=\footnotesize] {\isaf{extract\_map}} (gs);
  \node[font=\footnotesize\itshape, color=black!70, below=0.15cm of ads.south east, xshift=1.7cm, align=center]
    {blinding $\mapsto$ refinement,\; merge $\mapsto$ join};
\end{tikzpicture}
\caption{Coupling the state-preservation functor to the authenticated layer. The map \isaf{extract\_map} carries authenticated data to global states, sending the Merkle blinding order to the partial-view preorder \isaf{state\_refines} and the merge to \isaf{state\_join}. The soundness theorems below invoke the Merkle interface explicitly (not merely importing it). The coupling is instantiated on \isaf{ADS\_Functor}'s blindable-position functor and on a recursive model of the Canton transaction tree.}
\label{fig:ads}
\end{figure}

\subsection{The Authenticated-State Locale}
\label{sec:auth-state}

The \isaf{authenticated\_state} locale sits over a \isaf{merkle\_interface} (hash $h$, blinding order \isaf{bo}, merge $m$) together with an extraction map to global states, and assumes that extraction respects merging (into a \isaf{state\_join}), respects blinding (into a \isaf{state\_refines}), and yields only valid states. Its soundness theorem then transports authentication to the global-state level: if two authenticated data merge, their extracted states \emph{join} into a valid state that refines both. The join conjunct is the load-bearing one: it says the combined state is the least upper bound of the two partial views, carrying no chain that neither view authenticates, and the two refinement conjuncts are what that join gives on each side.

\needspace{9\baselineskip}
\begin{lstlisting}
theorem authenticated_preservation_soundness:
  assumes "m a b = Some ab"
    and "extract_map a = Some sa"
    and "extract_map b = Some sb"
  shows "∃sab. extract_map ab = Some sab
              ∧ valid_state sab
              ∧ state_join sa sb sab
              ∧ state_refines sa sab ∧ state_refines sb sab"
\end{lstlisting}

The proof uses the Merkle interface, it does not merely import it: it invokes \isaf{merge\_respects\_hashes} (equal hashes witness mergeability), \isaf{join} (the merge is a least upper bound), and \isaf{hash}. A companion theorem (\isaf{blinded\_view\_preserves\_validity}) discharges the need-to-know guarantee: a blinded, partial view is still a valid, consistency-respecting view. To keep these theorems off the empty class, the locale is instantiated concretely on the regulatory model (an authenticated hash over the consensus state, merge as revealed-chain union), with non-triviality witnesses.

\subsection{Categorical Laws of the Authenticated Layer}
\label{sec:auth-laws}

On this base the authenticated layer carries the category laws of the composite functor from the ADS blinding preorder to the cross-domain refinement preorder. Composition of blinding morphisms is preserved,

\begin{lstlisting}
theorem cdsp_ads_compose:
  assumes "bo a b" and "bo b c" and "extract_map c = Some sc"
  shows "bo a c ∧ (∃sa sb. extract_map a = Some sa
          ∧ extract_map b = Some sb
          ∧ valid_state sa ∧ valid_state sb ∧ valid_state sc
          ∧ state_refines sa sb ∧ state_refines sb sc
          ∧ state_refines sa sc)"
\end{lstlisting}

\noindent (the authenticated analogue of \isaf{preservation\_compose}), and the lifted merge is associative:

\needspace{8\baselineskip}
\begin{lstlisting}
theorem cdsp_ads_merge_assoc:
  assumes "m a b = Some ab" and "m ab c = Some abc"
  shows "∃bc. m b c = Some bc ∧ m a bc = Some abc"
\end{lstlisting}

Commutativity and idempotence descend directly from the Merkle interface laws; associativity is the one that needs rebracketing and so is recorded as a theorem. These laws are extended along whole blinding paths: every intermediate view on a path extracts to a valid state that refines the endpoint and commits to the same root hash (\isaf{sequence\_authenticity\_preservation}, \isaf{blinding\_path\_hash\_soundness}), and any revealed holding is genuinely present at the endpoint (\isaf{sequence\_inclusion\_integrity}). These are state-level (revealed-holding) statements; a separate development sharpens them to concrete Merkle inclusion paths, with a forgery-exclusion witness.

\subsection{Instantiation on Canton, and Its Scope Limit}
\label{sec:canton-instance}

The coupling is instantiated three times, at deliberately different fidelities, and the difference is part of the claim. \isaf{oss\_blindable} instantiates it on the blindable-position functor of \isaf{ADS\_Functor.ADS\_Construction}: the smallest carrier on which the authenticated laws can be exercised at all. \isaf{canton\_authenticated} instantiates it on a \emph{coarse} model of the Canton transaction: faithful at the top level, with the view structure collapsed, which suffices for the merge and blinding laws but cannot express nesting. \isaf{reg\_tx\_authenticated} instantiates it on a \emph{recursive} model, built over \isaf{ADS\_Functor.Canton\_Transaction\_Tree}'s rose-tree view structure, so that subview nesting and subview-level selective disclosure are faithful. The coarse instance is not superseded scaffolding: it isolates which of the guarantees survive without recursion, and the recursive one shows what recursion adds. Both instantiations transport the Merkle interface and the three coherence obligations through a constructor bijection, so no new axiom and no residual proof obligation arise; in particular, no axiom relates the opaque Canton datatypes to the regulatory model.

The recursive instance declares its fidelity boundary explicitly, and we reproduce it because it bounds the guarantee honestly. Two points are modelling choices rather than a one-to-one match with Canton's datatype. Leaf content is concrete \isaf{chain\_id}/\isaf{reg\_state} where Canton's leaves are opaque view data; this is recursive-faithful but content-abstracted. And the consensus \isaf{reg\_state} is modelled as a bare, non-independently-blindable field, whereas Canton's common metadata is itself an independently blindable position. This second choice is what structurally forces the single-consensus invariant---a view is reachable only by revealing the whole transaction, hence its consensus---which in turn is why validity holds with no assumption. The price is a scope limit: the case ``a view is revealed while the consensus is blinded'' is out of this model's scope; lifting it would require an independently blindable consensus and a fresh consistency argument, recorded as future work and as the residual fidelity check for the Canton authors, not closed here. That the single-consensus invariant is load-bearing rather than incidental is itself proved: a rogue state disagreeing across chains is exhibited as invalid, and the extraction never produces one.

\section{Proof Architecture, Genericity and Automation}
\label{sec:genericity}

Three things substantiate the claim that this is reusable infrastructure rather than a single formalization with generic-sounding names: a layered architecture in which each layer states what it fixes and what it leaves open, a proof-automation layer that discharges whole instances mechanically, and an instance drawn from outside the regulatory domain that uses only the generic theories. Two further subsections report what that architecture costs, in the form of where the proof effort landed once the layers were cut this way, and what we observed about the proof assistant itself.

\subsection{The Layered Architecture}
\label{sec:architecture}

The ten generic locales are not a flat pool. They stack into four layers, and the stacking is what makes the later results cheap to state (\Cref{tab:layers}).

\begin{table}[ht]
\centering
\caption{The four layers, what each fixes, and what an importer must supply.}
\label{tab:layers}
\small
\begin{tabular}{@{}P{2.5cm} P{4.2cm} P{4.6cm}@{}}
\toprule
\textbf{Layer} & \textbf{What it fixes} & \textbf{What an importer supplies} \\
\midrule
Generic
& State machines, preservation morphisms, roundtrips, $N$ domains; deterministic selection and fair scheduling
& Sets, a transition function, a terminal set; for morphisms, a state map and an action map \\
\addlinespace
Instance
& The regulatory vocabulary: five states, seven actions, the synchronization protocol
& Nothing; this layer is the application \\
\addlinespace
Compositional
& A guarded invariant, a windowed fairness source and a well-founded measure, yielding bounded convergence from an arbitrary carrier state
& A measure that strictly decreases on discharging steps outside the invariant \\
\addlinespace
Authenticated
& Merge and blinding lifted to global states, with the category laws of the composite
& A \isaf{merkle\_interface} and an extraction map satisfying three coherence obligations \\
\bottomrule
\end{tabular}
\end{table}

Two properties of the stack are worth stating because they are what a reader would otherwise have to reconstruct from the sources. First, the layers are separated by \emph{import}, not merely by naming: \isaf{Priority\_Resolution.thy} does not import \isaf{Regulatory\_Instance.thy} (\Cref{fig:import-graph}), and \isaf{External\_Instance.thy} imports neither. An accidental regulatory assumption in the generic layer would break those imports, so domain independence is enforced by the build rather than asserted in prose. Second, each layer's obligations are discharged once and reused: the convergence theorem of \Cref{sec:combined} is obtained by supplying \isaf{inconsistency\_pairs} as the measure to an already-proved generic theorem, not by re-running the convergence argument on global states.

\begin{figure}[!htb]
\centering
\begin{tikzpicture}[
  file/.style={rectangle, rounded corners=2pt, draw=black, fill=gray!5, minimum width=3.5cm, minimum height=0.7cm, font=\small\ttfamily},
  arr/.style={-{Stealth[length=5pt]}, thick},
  node distance=1.2cm and 2cm
]
  \node[file] (sp) {State\_Preservation.thy};
  \node[file, below=of sp] (ri) {Regulatory\_Instance.thy};
  \node[file, right=3cm of sp] (pr) {Priority\_Resolution.thy};
  \node[file, below=2.5cm of pr, xshift=-1cm] (dq) {DQuencer\_Instance.thy};

  \draw[arr] (sp) -- (ri);
  \draw[arr] (ri) -- (dq);
  \draw[arr] (pr) -- (dq);

  \draw[dashed, gray, thick] (pr) -- node[midway, above, sloped, font=\footnotesize\itshape, color=black!65] {no import} (ri);
\end{tikzpicture}
\caption{Import dependency graph. \texttt{Priority\_Resolution.thy} does not import \texttt{Regulatory\_Instance.thy} (dashed line), preserving domain independence. \texttt{DQuencer\_Instance.thy} imports both, serving as the join point.}
\label{fig:import-graph}
\end{figure}

\subsection{Where the Proof Effort Went}
\label{sec:effort}

The category laws of \Cref{sec:category-laws} are short. Counting each statement together with
its proof script, \isaf{preservation\_id} closes in 7 lines, \isaf{preservation\_assoc} in 25
and \isaf{preservation\_compose} in 42. Taken by themselves those numbers invite the reading
that the categorical layer is decoration. The measurement we want to report is the opposite
one, and it is a fact about the decomposition rather than about the subject: the laws are short
\emph{because} \isaf{state\_preservation} was cut so that its obligations are exactly the
naturality conditions, and once objects and morphisms are cut that way the category axioms have
no work left to do at that spot. \isaf{preservation\_id} is an unfolding followed by
\isa{simp}. \isaf{preservation\_assoc} obtains both bracketings as morphisms by two appeals to
\isaf{preservation\_compose} and then closes on \isa{comp_assoc}.

The work does not vanish under such a decomposition. It relocates to the points where a
concrete carrier has to meet the obligations the locale fixed, and in this development it
landed in three identifiable places (\Cref{tab:effort}).

\begin{table}[ht]
\centering
\caption{Statement-plus-proof length of representative results, counted in lines of the
released sources from the declaration keyword through the closing step of the proof. The
category laws are among the cheapest results in the development; the obligations of the
concrete carriers are the most expensive.}
\label{tab:effort}
\small
\begin{tabular}{@{}l r l@{}}
\toprule
\textbf{Result} & \textbf{Lines} & \textbf{What it is} \\
\midrule
\isaf{preservation\_id} & 7 & category law \\
\isaf{preservation\_assoc} & 25 & category law \\
\isaf{preservation\_compose} & 42 & category law \\
\addlinespace
\isaf{deg\_run\_collapse} & 21 & non-degeneracy of the degree tower \\
\quad \isaf{broadcast\_le\_overwrite} & 24 & lemma carrying its induction \\
\quad \isaf{deg\_run\_on\_broadcast} & 31 & lemma carrying its induction \\
\addlinespace
\isaf{reg\_chains\_blinding} & 30 & recursive lemma on the Canton tree \\
\isaf{reg\_chains\_merge} & 48 & recursive lemma on the Canton tree \\
\isaf{reg\_tx\_authenticated} & 130 & interpretation on the Canton carrier \\
\bottomrule
\end{tabular}
\end{table}

\paragraph{The authenticated interpretation.} The largest single proof block in the development
is the \isa{interpretation} \isaf{reg\_tx\_authenticated}, at 130 lines, three times the longest
category law. It states nothing of its own: every one of those lines discharges the three
coherence obligations of the \isaf{authenticated\_state} locale for the recursive Canton
transaction-tree carrier of \Cref{sec:canton-instance}. It is also shorter than it would
otherwise have been, by design. The three obligations were reduced in advance to two recursive
lemmas over the tree, \isaf{reg\_chains\_blinding} (30 lines) and \isaf{reg\_chains\_merge}
(48 lines), together with the state-level glue lemmas of \Cref{sec:glue}, so that the
recursive difficulty is isolated in two named results instead of being distributed through the
interpretation. Those two lemmas are where the development placed its own milestone boundary:
they were closed and reviewed before the interpretation was attempted.

\paragraph{The list merge underneath them.} Both recursive lemmas rest on a positional
characterization of the authenticated list merge, \isaf{merge\_list\_Cons}, whose proof is
6 lines and whose derivation was not proportionate to that: the transfer package does not fire
on this goal (\Cref{sec:tool-feedback}), so the lifted definition is unfolded directly and
closed on the representation equations. Six lines here record a route that had to be found
rather than a step that was easy.

\paragraph{The degree tower.} \isaf{deg\_run\_collapse} is the clearest case in the development
where a line count carries no information about difficulty. It is the result that separates
$F(k)$ from a vacuous fold over an index set, and \Cref{sec:degree-functors} states why its
proof is not a free-monoid triviality. Its own script is 21 lines, and it is 21 lines because
\isaf{broadcast\_le\_overwrite} (24 lines) and \isaf{deg\_run\_on\_broadcast} (31 lines) carry
the induction beneath it.

\paragraph{What the distribution measures.} Under a locale decomposition, proof length at a
given result measures how well the decomposition was chosen at that spot. It does not measure
how hard the mathematics is there, and the two come apart in both directions here: the category
laws are cheap because the layering was chosen to make them cheap, and \isaf{deg\_run\_collapse}
is cheap because two lemmas beneath it are not. The cost of the layering is visible in the
files: the three category laws occupy 74 lines of \isaf{Functor\_Laws.thy}, while
\isaf{Canton\_Bridge.thy}, which introduces no generic locale at all and exists to satisfy
obligations that other layers fixed, is 1,359 lines. A reader deciding where to spend review
effort on this artifact should read the interpretations and the recursive lemmas rather than
the laws.

\subsection{Reusable Proof Automation}
\label{sec:automation}

\isaf{Proof\_Automation.thy} provides two Eisbach methods that discharge the obligations of \isaf{state\_machine} and \isaf{state\_preservation} instances on finite, enumerated domains. Both unfold the locale predicate to atomic obligations and close each by simplifier-driven exhaustive case analysis, driven by three named theorem collections that an instance populates with its defining equations:

\begin{lstlisting}
method discharge_state_machine =
  (unfold_locales; (auto simp add: discharge_simps
     simp del: discharge_dels intro: discharge_intros
     split: option.splits if_splits))
\end{lstlisting}

Exercised against the development's own instances, the methods close \textbf{39 of 39} atomic obligations by pure search across four instances (a $6$-, an $11$-, an $11$-, and an $11$-obligation instance), with no registered interpretation to fall back on, and---the honesty point the sources state outright---``no obligation weakened or left to a manual fallback'': the discharged statements are the full locale predicates. Three further one-line restatements are discharged from existing registrations. Of the four search-closed instances, one is genuinely new (a forward escalation-scoped bridge with no manual counterpart, which the method constructs outright); the other three re-derive statements identical to their manual counterparts.

\paragraph{What the automation does not reach.} The reach is bounded, and the bound is structural rather than incidental. The methods close obligations that reduce to exhaustive case analysis over enumerated sets, which is exactly the obligation shape of \isaf{state\_machine} and \isaf{state\_preservation} on a finite, enumerated domain. Everything whose obligations need a genuine induction or an existential witness is proved by hand: the roundtrip conditions of \isaf{symmetric\_state\_preservation}, the consistency and isolation theorems of \isaf{multi\_domain\_preservation}, both liveness locales, the three coherence obligations of \isaf{authenticated\_state}, and every obligation of the compositional layer. Sequential preservation, the theorem that carries naturality from single transitions to action words, is an induction and is not automated either---the methods discharge the \emph{instances} of the locale from which it then follows for free. The honest summary is that the automation removes the boilerplate of admitting a new domain into the framework; it does not remove any of the proof work that produced the framework's theorems.

\subsection{Non-Vacuity: What Keeps the Theorems Off the Empty Class}
\label{sec:nonvacuity}

A locale-based development can be entirely correct and entirely empty. Four kinds of evidence are used here against that, and they are of increasing strength.

\emph{Interpretations.} Eighteen interpretation commands register concrete instances, and every generic locale is inhabited by at least one of them (\Cref{sec:bridge-instance}).

\emph{Ambient-hypothesis-free witnesses.} An interpretation performed inside a context establishes only relative consistency. Two witnesses, \isaf{singleton\_dq} and \isaf{singleton\_dq\_priority}, discharge the D-quencer host locale's assumptions with no ambient hypotheses at all, and \isaf{oss\_authenticated} inhabits the authenticated layer, so those locales are satisfiable outright.

\emph{Non-degenerate witnesses.} Satisfiability by a degenerate instance proves little: a one-node roster with no Byzantine node satisfies a Byzantine threshold trivially. The \isaf{bft\_quorum} interpretation is therefore taken at the tight bound, four nodes with one Byzantine node, where the progress obligation is real.

\emph{Deletion-sensitive witnesses.} The strongest of the four, because it fails loudly. Three regression witnesses exhibit a concrete counterexample to a guarantee if a guard is dropped: \isaf{guard\_is\_load\_bearing} (drop the DAML well-formedness guard and naturality breaks), \isaf{rogue\_join\_excluded} (drop the containment clause and the lifted merge admits a chain neither view authenticates), and \isaf{single\_consensus\_load\_bearing} (drop the single-consensus invariant and extraction can produce an invalid state). These do not merely show the hypotheses are satisfiable; they show the hypotheses are doing work.

\subsection{An Instance Outside the Regulatory Domain}
\label{sec:external}

To exhibit domain independence, the generic \emph{safety} layer is instantiated on a subject with no regulatory content: a TCP connection state machine (a subset of RFC~793, folded to a single-endpoint view) observed by a stateful connection tracker. The theory imports only the generic theories---not \isaf{Regulatory\_Instance.thy}---which is itself the evidence that the generic layer carries no hidden regulatory assumptions. Each obligation is closed by the generic method:

\begin{lstlisting}
theorem tcp_state_machine:
  "state_machine tcp_states tcp_events
                 tcp_transition tcp_terminal"
  by discharge_state_machine

theorem tcp_conntrack_preservation:
  "state_preservation
     tcp_states tracked_events tcp_transition tcp_terminal
     ct_states ct_events ct_transition ct_terminal
     tcp_to_ct tcp_to_ct_event"
  by discharge_preservation
\end{lstlisting}

The tracker record binds a peer-endpoint field to its phase tag exactly as the DAML permission record of \Cref{sec:bridge-instance} does, and the preservation morphism scopes its source actions to a strict subset (the tracker does not follow \isaf{RST}: an aborted connection is expired by timeout), reusing the same partial-action-vocabulary handle as the escalation instance. A non-vacuity witness runs a three-way handshake through both machines. This is a single instance chosen to \emph{exhibit} independence for the safety layer; it is a domain-independence witness, not a claim of domain coverage, and the liveness and authenticated layers are not exercised on it.

\subsection{Lessons for Other Mechanizations}
\label{sec:lessons}

Four things generalize past this subject matter, and we state them because each cost us a redesign.

\paragraph{Put the domain independence in the import graph.} A generic locale is only as generic as its imports allow. Keeping \isaf{Priority\_Resolution.thy} free of any regulatory import turned an unfalsifiable claim into a build-time check, and the external instance of \Cref{sec:external} is then evidence rather than assertion, because it compiles against the generic theories alone.

\paragraph{Move a delicate side condition into a sublocale rather than into the abstraction.} Priority injectivity holds for D-quencer messages only under a fixed-precision clock and a fixed node roster. Weakening the generic \isaf{priority\_system} locale to accommodate that would have contaminated every future importer with bounds that are not properties of priorities. Instantiating the generic locale on the key type instead, and recovering the message level inside two sublocales that carry the bounds explicitly, kept the abstraction intact (\Cref{sec:dquencer-instance}). The general rule: when an instance does not quite fit, adapt the instance's presentation, not the abstraction's assumptions.

\paragraph{Prefer a constructive witness to an absence of counterexamples.} Counterexample search inside a headless batch build reports nothing usable (\Cref{sec:tool-feedback}), and a bounded search that finds nothing is weak evidence in any case. Every non-vacuity claim here is instead carried by a concrete inhabitant, and every load-bearing claim by an instance that breaks when the guard is deleted (\Cref{sec:nonvacuity}). The witnesses cost more to write and are worth strictly more.

\paragraph{Let the formalization talk back to the design.} The distinction between best-effort and enforced mutual rollback entered the model because collapsing the two degrees left the ladder of \Cref{sec:tower} with no order to be monotone along; the degree hierarchy is four-valued in the product because it had to be four-valued for the natural transformations to compose. Mechanization is usually described as a check on a design that already exists. Here it was also a source of design constraints, and the constraints it produced were the ones a paper design had left implicit.

\subsection{Feedback on Isabelle/HOL}
\label{sec:tool-feedback}

The development was carried out in Isabelle/HOL 2025-2 and driven from a headless batch build,
so that every claim in this paper is reproduced by the same command a reader runs. We record
the points at which the tool rather than the subject decided how we proceeded. One is a limit
of what the logic lets us state; the rest are limits of what the tools report, or of what a
batch build lets the developer see. Each is given with what we were attempting, what stopped
us, and what we did instead.

\paragraph{A batch build has no channel back to the developer.} We wanted to precede each
non-vacuity argument with a counterexample search, and to inspect intermediate goals while
developing the recursive proofs of \Cref{sec:canton-instance}. In our configuration neither is
available under \isa{isabelle build}: the report from \isa{nitpick} is not surfaced in the build
log, and \isa{ML} \isa{writeln} output did not reach it either. What we did instead was to
provoke a deliberate proof failure on a goal we knew could not close, so that the goal state
would appear in the error message. That is the only route we found from a batch build back to
the developer, and it is plainly not one the system intends. At the level of results the
workaround was to abandon counterexample search in favour of constructive witnesses
(\Cref{sec:nonvacuity}), which are strictly stronger evidence, so this limitation improved the
artifact. The missing feature is a supported way to route diagnostic output, from \isa{nitpick}
and from \isa{ML} alike, into the log of a batch build.

\paragraph{Automation that declines without saying why.} Two standard tools stopped without a
diagnostic. The transfer package left the goal unchanged when we tried to characterize the
authenticated list merge positionwise through its lifted representation; it neither applied nor
reported an obstruction, and the output did not distinguish a goal outside its scope from an
incomplete relator setup on our side. We unfolded the lifted definition directly and closed on
the representation equations instead (\isaf{merge\_list\_Cons}), which also avoids a loop,
because unfolding the merge first leaves the recursion stopped at a variable tail. Separately,
a blanket \isa{split: option.splits} given to the simplifier diverges on the recursive
rose-tree structure rather than terminating in a failure, so those proofs have to be assembled
from single controlled unfoldings. The missing feature is the same in both cases: a report of
why a method did not apply, or of where a simplifier run is looping, in place of an unchanged
goal or a timeout.

\paragraph{An input-only abbreviation leaves two surface forms in play.} \isaf{hash\_discrete},
which the Canton carrier uses at every leaf, is declared in \isaf{ADS\_Functor} as an
\isa{abbreviation (input)} for \isa{id}. It is therefore expanded when a goal is stated and
never folded back when one is displayed, so what the goal shows is whatever the surrounding
simplification left behind, in some places \isa{id} and in others an explicit abstraction.
\isa{obtain} and \isa{simp} then fail to match a lemma against a goal it is equal to. The workaround generalizes the affected lemmas over an abstract function
variable and instantiates them at the point of use, so that no unfolding mismatch can arise:
\isaf{reg\_chains\_merge\_list} is stated over a general elementwise merge for this reason, and
not because the generality was wanted. This is a small thing that cost time out of proportion
to its size, and a more confluent normalization of eta-expanded constants across the tactic set
would remove it.

\paragraph{Whiskering is not statable in this encoding.} The tower of \Cref{sec:tower} encodes
an object as a record of a carrier set and a step function over one fixed HOL type of global
states, and a natural transformation as a function between global states. Vertical composition
is then a composition of HOL functions and is proved (\isaf{nt\_compose}). Horizontal
composition and whiskering are not merely unproved here but unstatable: they compose a
transformation with a functor whose source and target objects live in other types, and this
shallow encoding has no term denoting a functor as an action on the objects and morphisms of an
arbitrary category. Stating them would require a deep embedding, a datatype of categories and
functors carrying its own well-formedness conditions, which would put the tower in a different
idiom from the rest of the development, where objects are HOL types and morphisms are HOL
functions, and would forfeit the direct use of the \isaf{ADS\_Functor} combinators that
\Cref{sec:ads} applies. We record the omission as a consequence of that trade rather than as a
gap in the tower, and \Cref{sec:nat-tower} claims only vertical composition accordingly. A
reader who needs the 2-categorical structure should expect to pay for a deep embedding.

\paragraph{The session database is not crash-safe.} Interrupting \isa{isabelle build} while it
is writing leaves the session database in a state that fails the next build with a primary-key
conflict and exit code 2. The failure is reported in the same shape as a failed proof, so the
first diagnosis is the wrong one; recovery is to delete the session's heap log database by hand
and rebuild. On a development driven from scripted builds, where interruption is routine, a
crash-safe write or an exit status distinguishable from a proof failure would remove a recurring
misdiagnosis.

\paragraph{Two frictions that were ours.} We separate these because they are not defects of the
system. A locale parameter we named \isa{extract} collides with a command keyword and had to be
renamed; the collision is correct behaviour, and only the parse error, which does not name the
keyword it collided with, cost us anything. And a forced rebuild of the full session takes
minutes because the import closure is large, which is a property of our dependencies rather
than of the tool. The second is worth recording anyway, because the way around it was the
single change that most improved throughput here: cache the main session once, then develop
each proof in a one-theory session that imports it, which brings the edit-and-check cycle down
to seconds. We found that pattern nowhere in the documentation.

\section{Related Work}
\label{sec:related}

\subsection{Formal Verification of Blockchain Protocols}
\label{sec:related-blockchain}

\paragraph{Merkle Functor pattern.} Lochbihler and Mari\'{c}~\cite{lochbihler2020} abstracted Canton's authenticated data structures as Merkle Functors in Isabelle/HOL (FMBC 2020, AFP entry). Their work focuses on data structure integrity (authentication, inclusion proofs) and does not address cross-domain state transition preservation or Byzantine environments. Canton's formal verification roadmap~\cite{digitalasset2023} lists Byzantine environment verification as ``future work.'' This work uses the same tool (Isabelle/HOL) and methodology (locale-based modularization) but targets a different abstraction axis: state transition preservation rather than data structure integrity. Rather than remaining parallel, we couple to their entry directly (\Cref{sec:ads}), lifting its merge and blinding operations to the global-state level, and build a tower of degree-indexed functors with natural transformations---structure that, to our knowledge, their entry does not develop.

\paragraph{Verified authenticated data structures.} Beyond \isaf{ADS\_Functor}, Gregersen, Agarwal and Tassarotti~\cite{gregersen2025} give a formal proof of security and correctness for a library that generates authenticated versions of data structures automatically, on the basis of a new relational separation logic for programs using collision-resistant hash functions (CCS 2025). The two developments sit at different levels and answer different questions. Theirs is about the generator: whether the authenticated version a library produces is sound, and whether hand-written optimizations can be soundly linked against generated code. Ours takes an authenticated data structure as given and asks what survives when its merge and blinding are lifted to a state shared across domains (\Cref{sec:ads}). Neither result implies the other, and a development that wanted both would want their soundness beneath our lifting.

\paragraph{Velisarios.} Rahli et al.~\cite{rahli2018} verified PBFT's agreement (safety) in Coq (ESOP 2018). Their Logic of Events--based framework performed implementation-level verification of PBFT, but liveness was out of scope.

\paragraph{HotStuff/LibraBFT.} Carr et al.~\cite{carr2022} verified HotStuff's safety in Agda (NFM 2022), stating that liveness properties ``would be proved for specific implementations, not for the abstract model.''

\subsection{BFT Liveness Verification}
\label{sec:related-liveness}

Byzantine consensus liveness has been verified across various tools. TLA+ supports model checking for finite state spaces. Coq-based Velisarios~\cite{rahli2018} verified PBFT safety only. In Isabelle/HOL, the Heard-Of model~\cite{charronbost2009} has been used to verify round-based consensus algorithms, and Wanner et al.~\cite{wanner2020} verified both safety and liveness of a log replication protocol (SRDS 2020) using the Heard-Of model in Isabelle/HOL. Their work targets a leaderless protocol for secure logging, not leader-based blockchain consensus. Losa and Dodds~\cite{losa2020} verified both safety and liveness of the Stellar Consensus Protocol (FMBC 2020) in a combination of Ivy and Isabelle/HOL: the protocol and its temporal properties are established in Ivy's first-order setting, with Isabelle/HOL validating the first-order encoding against a more direct higher-order model. Bertrand et al.~\cite{bertrand2022} verified safety and liveness of the Red Belly Blockchain consensus holistically for any $n$ and $f < n/3$, decomposing the algorithm into threshold automata and model checking LTL properties (DISC 2022). At the chain level, P\^{\i}rlea and Sergey~\cite{pirlea2018} mechanized blockchain consensus in Coq, and Jones and Marmsoler~\cite{jones2024} verified the common-prefix property of proof-of-work consensus in Isabelle/HOL; both address consensus- or chain-level properties rather than cross-domain state semantics.

Compositionality has also been pursued directly. Zhao, P\^{\i}rlea et al.~\cite{zhao2024} build Bythos, a Coq framework in which Byzantine protocols are implemented, composed and verified, with an OCaml implementation obtained by extraction (CCS 2024). It is the closest work to ours on the compositional axis and it is ahead of us on one that we do not attempt: its verified protocols reach running code. The composition it performs is between protocols in one system, whereas ours is between the same protocol's effect on several domains, and the objects composed are morphisms rather than protocol layers. The comparison is worth stating plainly because it marks the distance to a refinement we declare out of scope (\Cref{sec:future-work}).

This work differs in two respects. First, none of the mechanized BFT liveness results above is carried out in Isabelle/HOL alone: the Heard-Of development targets round-based, leaderless protocols, the Stellar proof leans on Ivy's first-order automation, and the Red Belly proof proceeds by model checking. This work verifies liveness of leader-based BFT consensus (VRF leader election, epoch-based progression) in Isabelle/HOL alone. Second, existing work targets generic consensus properties (agreement, validity, termination), while this work verifies regulatory-specific properties: deterministic resolution by regulatory authority priority and starvation freedom for regulatory requests. To the best of our knowledge, no prior work has formally verified such domain-specific consensus properties.

\subsection{Cross-Chain State Consistency}
\label{sec:related-cross-chain}

Formal work on cross-chain state consistency is comparatively thin, and what exists targets bridge protocol security rather than structural preservation of state transitions. The closest mechanized neighbour is Mari\'{c}, Scholz and Suboti\'{c}~\cite{maric2025}, who model a fail-safe cross-chain bridge in Isabelle/HOL and verify it against formally stated security requirements by stepwise refinement from an abstract model (FMBC 2025). The comparison is instructive in both directions. Their subject is a bridge---a two-party protocol whose correctness is the preservation of a token invariant across a transfer, including under failure---and their method is refinement, an inherently vertical relation between one abstract model and one concrete one. Ours is horizontal: the same state machine replicated over an arbitrary finite domain set, related by morphisms that compose, with the guarantee being that a regulatory transition is reflected identically everywhere. Neither subsumes the other, and a bridge is in fact a natural future consumer of the morphism layer developed here. Beyond bridges, Bertrand et al.~\cite{bertrand2022} and P\^{\i}rlea and Sergey~\cite{pirlea2018} address consensus- and chain-level agreement, not the cross-domain semantics of an individual asset's regulatory state. To the best of our knowledge, no prior work unifies bidirectional roundtrip guarantees, $N$-domain consistency, and per-asset isolation in a single locale hierarchy.

\subsection{Positioning, and Two Tooling Choices Not Taken}
\label{sec:positioning}

\paragraph{Which claim is which.} This paper makes claims of two different kinds, and they should be judged separately. The \emph{methodological} claims are that cross-domain state preservation admits a functorial organization, that a tower of degree-indexed functors with natural transformations sits above it, and that the composite with an authenticated data structure again satisfies the category laws. Each is a theorem, none requires a new proof technique, and their novelty is a matter of what has been mechanized before. The \emph{application} claim is different in kind: that the object being formalized---the atomicity of regulatory effect across heterogeneous domains, together with the terminal-faithfulness of the recovery that restores consistency---is one that formal methods have not previously been pointed at. That claim is not made stronger by the proofs being hard; it is made stronger by the object being real, and the evidence for that has two parts, one observed and one structural. We state them separately because they carry different weight.

The observed part is that enforcement operations are deployed and their legal effect is left undefined by the specifications that define them. ERC-3643~\cite{erc3643} specifies freezing and forced transfer for regulated security tokens; ERC-7943~\cite{erc7943} specifies a forced-transfer interface for real-world assets. Neither assigns a legal effect to the operation it defines, and ERC-7943 says so deliberately, choosing motive-neutral names precisely because names such as confiscation or recovery would commit the standard to one legal reading. The standard-by-standard analysis, including the older security-token standards, is in~\cite{kim2026rcp}.

The structural part is that none of these standards provides atomicity for an enforcement action across more than one ledger. This is an argument from the absence of a mechanism, not from an incident report: we do not claim to have observed a confiscation that took effect on one domain and not another. What we claim is narrower and, for a formal-methods audience, more to the point---nothing in the deployed specifications rules that state out, so it is reachable rather than merely unlucky. \Cref{sec:safety} is the statement that within an atomic model it is not reachable, and \Cref{sec:discussion} is explicit about the distance between that model and a network.

We separate the methodological and application claims because a reviewer who discounts the first should still be able to weigh the second.

\paragraph{Why not a general category-theory library.} Isabelle/HOL has mature category-theory developments in the AFP, notably Stark's \isaf{Category3}~\cite{stark2016}, which defines categories in object-free style and gives functors and natural transformations as functions on arrows. We do not instantiate them, and the reason is not oversight. Our objects are locale \emph{predicates} over concrete carriers, and the morphisms are pairs of ordinary functions constrained by naturality; the category laws are consequently three short theorems about locale predicates (\Cref{sec:category-laws}). Routing them through a general library would require exhibiting our state machines as arrows in an object-free category, which buys the general theory---limits, adjunctions, the Yoneda lemma---that nothing here consumes, at the cost of an encoding layer between every application theorem and its statement. The trade is worth revisiting the moment a result is wanted that the general theory supplies and we would otherwise have to prove; until then the direct statements are the ones a reader can check against the sources.

\paragraph{Why the probabilistic layer is deterministic.} The \isaf{fair\_leader} assumption abstracts a VRF-based leader election, and the argument that the abstraction is reasonable---that $k$ consecutive Byzantine leaders have probability $(f/n)^k$---is prose, deliberately outside the proof (\Cref{sec:assume-guarantee}). Isabelle/HOL can carry that argument: CryptHOL~\cite{basin2020crypthol} is built for exactly this kind of game-based probabilistic reasoning. We keep the layer deterministic anyway. Formalizing the election would change what the development is about, from cross-domain state semantics to randomized leader election, and it would not strengthen any theorem stated here: every result downstream of \isaf{fair\_leader} consumes the deterministic bound, not the distribution behind it. The honest statement of the boundary is that the probabilistic backing is cited, not mechanized, and that a CryptHOL treatment of the election is a separable piece of work whose interface to this one is already fixed by the \isaf{fair\_leader\_system} locale.

\section{Discussion and Limitations}
\label{sec:discussion}

\subsection{Reusability of Generic Locales}
\label{sec:reusability}

The reusability of the ten generic locales is a primary contribution. Each locale is domain-independent and instantiated via Isabelle/HOL's \isa{interpretation} mechanism. The transplant targets below are structural analogies, not additional formalizations: none is verified here.

\paragraph{priority\_system.} A potential transplant target is any setting requiring deterministic selection under a total order with injective keys, such as MEV auction bundle priority selection, distributed task scheduler job ordering, or multichain bridge relay message ordering.

\paragraph{fair\_leader\_system.} A potential transplant target is starvation-freedom verification under periodic honest scheduling, such as leader-based BFT consensus (Tendermint, HotStuff) or round-robin scheduler fairness.

\paragraph{state\_preservation + multi\_domain\_preservation.} A potential transplant target is any structure-preserving synchronization between representations, such as bidirectional database replication, distributed cache consistency, or state-channel on-chain/off-chain synchronization.

\paragraph{converging\_composition.} A potential transplant target is any bounded self-stabilization argument that composes a guarded invariant with a windowed fairness source and a well-founded progress measure, such as anti-entropy replication, distributed cache reconciliation, or eventual-consistency convergence bounds.

\subsection{Model Assumptions and Implementation Gaps}
\label{sec:gaps}

\begin{table}[t]
\centering
\caption{Model assumptions and implementation gap mitigation.}
\label{tab:gaps}
\small
\begin{tabular}{@{}P{3cm} P{3.5cm} P{4cm}@{}}
\toprule
\textbf{Model Assumption} & \textbf{Implementation Reality} & \textbf{Gap Mitigation} \\
\midrule
Atomic sync execution & Network delay, partial failure & Preemptive lock guard; contention-time locking deferred (future work) \\
Honest nodes (Prop.~1) & Byzantine nodes possible & Liveness under $f<n/3$ (Property~2); safety--liveness convergence in the compositional layer \\
\isaf{fair\_leader} & VRF randomness & VRF + forced leader rotation \\
Closed system & Dynamic request arrival & Open system model (future work) \\
Instant lock acquisition & Distributed lock contention & Lock queue; contention-time locking deferred (future work) \\
\bottomrule
\end{tabular}
\end{table}

Formal refinement between the model and implementation code (Rust/Solidity) is unproven. This gap is managed through a \isaf{FORMAL\_MODEL\_MAPPING.md} document~\cite{oraclizer2026github} that explicitly maps each model element to its implementation counterpart.

\subsection{Scope Boundary and Future Work}
\label{sec:future-work}

It is easier to read a formalization generously than accurately, so we state the boundary as a list of things this paper does \emph{not} establish.

\begin{itemize}[leftmargin=*]
\item \textbf{Not concurrency.} Synchronization is atomic. There is no interleaving, no message delay, no redelivery or reordering, and therefore no lock contention; deadlock-freedom is not a theorem here because contention is not in the model.
\item \textbf{Not network-level liveness.} Convergence and starvation freedom hold \emph{under} the \isaf{fair\_leader} assumption. The threshold $n \geq 3f+1$ is shown to make that assumption inhabitable; it is not shown to make it true of any network. Byzantine behaviour is a cardinality-bounded tag, not an adversary that equivocates or forges.
\item \textbf{Not an open system.} Requests do not arrive dynamically; the \isaf{non\_honest\_bounded} assumption closes the system.
\item \textbf{Not a refinement to code.} No formal relation is established between these models and the Rust or Solidity implementation. \isaf{FORMAL\_MODEL\_MAPPING.md}~\cite{oraclizer2026github} records the intended correspondence in prose, which is documentation, not proof.
\item \textbf{Not a separation result.} The degree tower is shown to occupy structure that, to our knowledge, \isaf{ADS\_Functor} does not develop. That is a statement about the literature, not a theorem that the tower is strictly beyond it.
\item \textbf{Not full Canton fidelity.} The recursive instance is recursive-faithful but content-abstracted, and its consensus field is not independently blindable, so a view revealed while its consensus is blinded is outside the model. Lifting that limit needs an independently blindable consensus and a fresh consistency argument.
\item \textbf{Not domain coverage.} One instance outside the regulatory domain witnesses that the generic safety layer carries no hidden regulatory assumption. One witness is evidence of independence, not a survey of applicability, and the liveness and authenticated layers have no external instance at all.
\end{itemize}

Each of these is a place where the present development stops rather than a defect in what it proves, and each names the work that would move it.

\subsection{Open Questions}
\label{sec:open-questions}

\begin{enumerate}[leftmargin=*]
\item Extension to a partially synchronous network model with message delays, redelivery, and reordering.
\item Consistency guarantees under dynamic domain topologies (chains added/removed at runtime).
\item Starvation freedom in an open system with dynamic request arrival.
\item Formal refinement between the model and implementation code.
\item A direction-of-reconciliation semantics: which record is authoritative when the recovery path must choose between disagreeing histories, beyond the terminal-faithfulness constraints established here.
\end{enumerate}

\section{Conclusion}
\label{sec:conclusion}

This paper presented a mechanized theory, in Isabelle/HOL, of regulatory action atomicity across heterogeneous domains, organized around a cross-domain state-preservation functor and carried out across ten theory files that build without \isa{sorry} or \isa{oops}.

\textbf{The functor.} State machines and their structure-preserving synchronization maps form a category on which the construction satisfies the identity, composition, and associativity laws as theorems, with each transition system a functor on its action words.

\textbf{Safety and liveness.} Bidirectional, multi-domain state preservation, with terminal states preserved so that regulatory finality survives synchronization; determinism and starvation freedom under $f < n/3$ Byzantine faults, with the threshold $n \geq 3f+1$ shown to make the fair-leader assumption inhabitable rather than vacuous.

\textbf{Convergence.} From an arbitrary unlocked configuration, with no initial-consistency assumption, synchronization converges to a valid state within a bounded number of steps under a fair-leader assumption, along a terminal-faithful recovery path that neither manufactures nor erases confiscations.

\textbf{Hierarchy and coupling.} A tower of synchronization-degree functors connected by natural transformations closed under composition, with a one-directional degree monotonicity; and a coupling to Lochbihler and Mari\'{c}'s authenticated data structure, instantiated on the Canton transaction tree with a declared consensus-scope limit.

The reusable locales, discharge methods, and an external instance support reuse beyond the regulatory domain. The synchronization model is atomic; its lift to a partially synchronous network is left to subsequent work. All artifacts are publicly available on GitHub as a versioned Isabelle session release~\cite{oraclizer2026github,kim2026artifact}.

\bibliographystyle{plain}

\appendix
\clearpage
\section{Proof Artifact Summary}
\label{app:artifacts}

\begin{table}[H]
\centering
\caption{Proof artifact summary.}
\label{tab:artifacts}
\small
\begin{tabular}{@{}l r l l@{}}
\toprule
\textbf{File} & \textbf{Lines} & \textbf{Role} & \textbf{Key Theorems} \\
\midrule
State\_Preservation.thy & 455 & Safety generic (4 locales) & \isaf{sequential\_preservation}, \\
& & & \isaf{cross\_domain\_consistency} \\
\addlinespace
Regulatory\_Instance.thy & 1,598 & Safety instance & \isaf{regulatory\_homomorphism}, \\
& & & \isaf{valid\_state\_preservation} \\
\addlinespace
Priority\_Resolution.thy & 354 & Liveness generic (2 locales) & \isaf{select\_highest\_deterministic}, \\
& & & \isaf{eventual\_completion} \\
\addlinespace
DQuencer\_Instance.thy & 651 & Liveness instance & \isaf{dq\_priority}, \\
& & & \isaf{conditional\_safety\_preservation} \\
\addlinespace
Functor\_Laws.thy & 2,116 & Functor + ADS + convergence & \isaf{preservation\_compose}, \\
& & & \isaf{authenticated\_preservation\_soundness}, \\
& & & \isaf{oraclizer\_guarded\_bounded\_convergence}, \\
& & & \isaf{recovery\_no\_fresh\_terminal} \\
\addlinespace
Composition.thy & 443 & Guarded convergence (generic) & \isaf{bounded\_convergence\_from\_arbitrary} \\
\addlinespace
Hierarchy.thy & 1,061 & Degree functor tower & \isaf{degree\_forget\_natural\_run}, \\
& & & \isaf{nt\_compose}, \isaf{no\_downward\_safety} \\
\addlinespace
Canton\_Bridge.thy & 1,359 & Authenticated layer + Canton & \isaf{cdsp\_ads\_compose}, \\
& & & \isaf{cdsp\_ads\_merge\_assoc} \\
\addlinespace
External\_Instance.thy & 315 & Domain-independence witness & \isaf{tcp\_conntrack\_preservation} \\
\addlinespace
Proof\_Automation.thy & 94 & Eisbach discharge methods & \isaf{discharge\_state\_machine}, \\
& & & \isaf{discharge\_preservation} \\
\midrule
\textbf{Total} & \textbf{8,446} & & \\
\bottomrule
\end{tabular}
\end{table}

\noindent The four base theories (top) carry the safety and liveness substrate; the six further theories develop the functor and its category laws, guarded bounded convergence, the synchronization-degree tower, the authenticated-data-structure coupling, an external instance, and the proof-automation layer. The dependency structure of the four base theories, on which the compositional theories sit, is shown in \Cref{fig:import-graph}.

\paragraph{Build and availability.} All files build in Isabelle/HOL 2025-2 without \isa{sorry}, \isa{oops}, \isa{axiomatization} or an oracle. The development is distributed as a versioned session release in the public repository~\cite{oraclizer2026github}, accompanied by a manifest that records the Isabelle version, the document engine, a hash of the source tree, and the exact build command, so that a reader can reproduce the check rather than take the claim on trust~\cite{kim2026artifact}. The session depends on one external development, Lochbihler and Mari\'{c}'s \isaf{ADS\_Functor}~\cite{lochbihler2020}, and is cited above by its repository release.

\end{document}